\documentclass[a4paper,11pt]{article}
\pdfoutput=1 

\usepackage{jcappub} 

\usepackage[T1]{fontenc} 
\usepackage[utf8]{inputenc}
\usepackage{dsfont}
\usepackage{enumerate}

\def\be{\begin{equation}}
\def\ee{\end{equation}}
\def\bea{\begin{eqnarray}}
\def\eea{\end{eqnarray}}
\def\ba#1\ea{\begin{align}#1\end{align}}

\def\dd{\mathrm{d}}
\def\xx{\mathbf{x}}
\def\kk{\mathbf{k}}
\def\hk{\hat{k}}
\def\hn{\hat{n}}

\def\mr{\mathrm}
\def\Cov{\mathrm{Cov}}
\def\Var{\mathrm{Var}}
\def\lbra{\left\langle}
\def\rbra{\right\rangle}
\def\nbargal{\ensuremath{\overline{n}_\mathrm{gal}}}

\def\Msun{M_\odot}

\def\mnras{MNRAS}

\usepackage{color}

\title{\boldmath  Combining cluster number counts and galaxy clustering}

\author[a]{Fabien Lacasa,}
\author[a]{Rogerio Rosenfeld}

\affiliation[a]{ICTP South American Institute for Fundamental Research\\
                Instituto de F\'{\i}sica Te\'orica, Universidade Estadual Paulista, S\~ao Paulo, Brazil}

\emailAdd{fabien@ift.unesp.br}
\emailAdd{rosenfel@ift.unesp.br}

\abstract{
The abundance of clusters and the clustering of galaxies are two of the important cosmological probes for current and future large scale surveys of galaxies, such as the Dark Energy Survey. In order to combine them one has to account for the fact that they are not independent quantities, since they probe the same density field.
It is important to develop a good understanding of their correlation in order to extract parameter constraints.
We present a detailed modelling of the joint covariance matrix between cluster number counts and the galaxy angular power spectrum. We employ the framework of the halo model complemented by a Halo Occupation Distribution model (HOD). We demonstrate the importance of accounting for non-Gaussianity to produce accurate covariance predictions. Indeed, we show that the non-Gaussian covariance becomes dominant at small scales, low redshifts or high cluster masses. We discuss in particular the case of the super-sample covariance (SSC), including the effects of galaxy shot-noise, halo second order bias and non-local bias. We demonstrate that the SSC obeys mathematical inequalities and positivity.
Using the joint covariance matrix and a Fisher matrix methodology, we examine the prospects of combining these two probes to constrain cosmological and HOD parameters. We find that the combination indeed results in noticeably better constraints, with improvements of order 20\% on cosmological parameters compared to the best single probe, and even greater improvement on HOD parameters, with reduction of error bars by a factor 1.4-4.8. This happens in particular because the cross-covariance introduces a synergy between the probes on small scales. We conclude that accounting for non-Gaussian effects is required for the joint analysis of these observables in galaxy surveys. 
}

\begin{document}
{\center \today}
\maketitle

\flushbottom


\section{Introduction}\label{Sect:intro}

The large scale structure of the universe has a long history of being a useful probe of both global cosmological parameters and the formation of structures in our universe. This is the motivation for carrying out large galaxy surveys, such as the Dark Energy Survey (DES\footnote{\url{http://www.darkenergysurvey.org}}), which will use different observational probes, such as weak gravitational lensing, correlations in the distribution of galaxies, statistics of galaxy clusters, and type Ia supernovae, in order to improve our understanding of background cosmology and the growth of structures. For example the large scale distribution of matter is sensitive to the presence and specific properties of dark energy, see e.g. \cite{Frieman:2008sn} for a review. On smaller scales however, the clustering is dependent on galaxy formation models and halo physics. 

Different observables have different sensitivities to cosmological parameters, and combining them can help to break degeneracies present when only single probes are used. However these probes are in principle not independent. Thus in order to estimate parameter constraints and their accuracy from these large surveys, it is necessary to have a careful understanding of the statistical properties of the probes used, including their cross-covariance.

In this work we will be interested in the combination of two probes, namely the galaxy angular clustering and the abundances of clusters. Indeed cluster counts can be used to test cosmological models through both volume effects and the growth rate \cite{Hu:2002we,Majumdar:2003mw,Lima:2005tt}, e.g. different dark energy models affect differently the formation of dark matter haloes \cite{Liberato:2006un,Abramo2007}.
Likewise, measurements of the galaxy correlation function can be used to estimate cosmological parameters and the formation of structures, see e.g. \cite{deSimoni:2013oqa} for a recent example with SDSS observations. 

Cluster counts and galaxy-galaxy correlations both probe the same density field, whose non-linearity couples different Fourier modes. Besides, some of the galaxies probed will be living in the same halos probed by the number counts. Through both these effects, we expect the two probes to be importantly correlated, in particular on small scales, where non-Gaussian effects are more relevant. Thus in order to determine the statistical power of combining cluster counts and 2-point galaxy clustering, we need to model the covariance of these observables as well as their cross-covariance.
We will do so in the context of the halo model framework to describe the non-linear density field (see, e.g., \cite{CooraySheth2002}). The auto-covariance of cluster counts and of the 2-point galaxy correlation function are described by 2-point and 4-point correlation functions respectively, whereas the cross-covariance is related to a 3-point correlation function.

The necessary ingredients for the modelling are the halo mass function, the halo density profile and a model to describe how galaxies populate dark matter halos. We will use the so-called Tinker mass function~\cite{Tinker2008}, the Navarro-Frenk-White (NFW) density profile~\cite{Navarro:1996gj} and a halo occupation distribution (HOD) model from~\cite{Seljak:2000gq,Scoccimarro:2000gm}. In addition to the usual cosmological parameters, we will study the sensitivity of these probes to the dark energy equation of state (assuming a flat wCDM model) and to HOD parameters.

We build on previous studies in the literature. For example the covariance of the matter power spectrum was predicted with the halo model by Cooray and Hu~\cite{Cooray:2000ry} and compared favourably with simulations. Takada and Hu~\cite{Takada:2013wfa} emphasized the importance of the so-called super-sample covariance in the matter power spectrum, due to long wavelength modes outside the volume of the survey. In this article we will derive and discuss the super-sample covariance of our observables. Cluster counts have already been shown to be a powerful probe for cosmological constraints, even more so when combined with other probes. In particular Takada and Bridle~\cite{Takada:2007fq} showed that it mitigates the super-sample covariance of the cosmic shear power spectrum. This was later generalized by \cite{Schaan:2014cpa} to 3D $n$-point correlation functions with a joint likelihood. Cluster counts were also combined with the cluster power spectrum in \cite{Mana2013}. Moreover, Hutsi showed that the cluster-galaxy correlation contains a significant amount of information for cosmological and HOD constraints \cite{Hutsi2008}. Finally for photometric surveys, a framework was developed to analyse projected angular probes of weak lensing and the galaxy distribution in \cite{Krause:2016jvl}.

In this article, we will develop a unified framework to treat the cluster number counts and the galaxy-galaxy angular power spectrum in the full-sky case. We pay particular attention to the non-linear effects at small scales, to a careful modelling of the cross-covariance of these probes, and deriving and discussing the super-sample covariance for our observables.

Our goal can be easily stated as follows. We will consider grouping the data from cluster counts and angular power spectrum into a single data vector $X$:
\be
X = \begin{pmatrix} N_\mr{cl} \\ C_\ell^\mr{gal}\end{pmatrix}
\ee
where $ N_\mr{cl}$ is the number of clusters in a given mass and redshift bin and $C_\ell^\mr{gal}$ is the angular power spectrum of galaxies in a given redshift bin (the dependence on mass and redshift are kept implicit for simplicity). We want to model the joint covariance matrix
\be
\Cov(\hat{X},\hat{X}) = 
\begin{pmatrix}
\Cov(\hat{N}_\mr{cl}, \hat{N}_\mr{cl}) & \Cov(\hat{C}_\ell^\mr{gal} , \hat{N}_\mr{cl}) \\
\Cov(\hat{N}_\mr{cl}, \hat{C}_\ell^\mr{gal}) & \Cov(\hat{C}_\ell^\mr{gal}, \hat{C}_\ell^\mr{gal})
\end{pmatrix}
\label{Eq:Cov}
\ee
This covariance matrix will be used to study the statistical power of these observables in estimating parameters, using a Fisher matrix approach. In particular we will investigate the relevance of accounting for the cross-covariance represented by the off-diagonal terms in the covariance matrix eq.~(\ref{Eq:Cov}).

The cross-covariance is related to a bispectrum involving a halo-galaxy-galaxy correlator. This bispectrum is modelled in detail, with contributions arising from perturbation theory, the halo model and HOD. In particular we include the effects of second order halo bias as well as non-local bias and galaxy shot noise. We are able to show explicitly how the super-sample covariance appears naturally in the bispectrum, in a framework that incorporates it in a systematic way.

This article is organized as follows. In the next section we set the notations and conventions that will be used throughout the article.
In section \ref{Sect:counts} we describe the cluster number counts and its covariance.
Next in section \ref{Sect:galspec} we model the galaxy angular power spectrum. In section \ref{Sect:crosscov} we describe the modelling of the cross-covariance between cluster counts and galaxy angular power spectrum. In section \ref{Sect:SSC} we propose a unified framework to treat the super-sample covariance among different probes. In section \ref{Sect:covgal} we discuss the covariance of the galaxy angular power spectrum. In section \ref{Sect:jointcov} we present the joint covariance and perform a Fisher matrix analysis of the constraints on cosmological and HOD parameters, respectively on large and small scales. Finally, we conclude in section \ref{Sect:conclusion}. Several appendices are presented with details of the computations and proofs of assertions from the main text.

\section{Notations, inputs and conventions}\label{Sect:notations}

As stated in the Introduction, we will base our analysis on the halo model framework.
In this framework, all galaxies are located in halos and one can write the number density of halos as:
\be \label{Eq:halodensity-sumi}
n_h(\xx) = \sum_i \delta^3(\xx- \xx_i) 
\ee
where the index i denotes halos, with center located at $\xx_i$ and mass $M_i$.

In particular the halo mass function, giving the comoving number density of halos per unit mass is given by the ensemble average \cite{CooraySheth2002}: 
\be 
\frac{d n_h}{dM} = \lbra {\cal N}_{cl}(\xx | M,z) \rbra, 
\ee
where
\ba
{\cal N}_{cl}(\xx | M,z) = \sum_i \delta^3(\xx-\xx_i) \delta(M-M_i).
\ea
Galaxies inhabit the dark matter halos and their number density can thus be written as: 
\be \label{Eq:galdensity-sumij}
n_\mr{gal}(\xx) = \sum_i \sum_{j=1}^{N_\mr{gal}(i)}  \delta^3(\xx- \xx_j),
\ee
where the index $j$ refers to galaxies in a given halo $i$ with $N_\mr{gal}(i)$ galaxies.
We further assume that within a halo, the galaxies are distributed independently following the normalized halo density profile
\be 
u(\xx-\xx_i | M_i) = \frac{1}{M_i} \rho(\xx- \xx_i | M_i) .
\ee
We will adopt the spherically symmetric NFW profile~\cite{Navarro:1996gj}
\be 
u(r|M_i) = \frac{1}{M}\frac{\rho_s}{r/r_s ( 1 + r/r_s)^2},
\ee
with $r_s$ the scale radius and $\rho_s$ is the scale density. They are specified as in, e.g.  \cite{Bullock:1999he,Komatsu:2002wc}.\\
The normalization enforces in particular that the Fourier transform $u(k|M,z)$ obeys
\be
u(k|M,z) \rightarrow 1 \quad \mr{when} \quad k \rightarrow 0.
\ee

For the mass function, we use the one from Tinker et al.~\cite{Tinker2008}, which is a fit to the number density of halos obtained from  simulations in cold dark matter scenarios.
The halo bias relating fluctuations in the halo density field $\delta_h$ and those  of the dark matter field  $\delta_m$ are parametrized via the local bias ansatz:
\be 
\delta_h = \sum_i b_i(M,z) \left(\delta_m^i - \lbra\delta_m^i\rbra\right),
\ee
where $\langle \delta_h \rangle = 0$ by construction and the biases are consistently related to the chosen mass function. 
We will use both the first~\cite{Tinker:2010my} and second order bias~\cite{Hoffmann:2015mma} in this work,  neglecting higher order terms.

Galaxies are introduced using a HOD model which specifies how galaxies populate halos.
In this model, the probability distribution of galaxies is a function of the halo mass only and is composed of two populations:
a  central galaxy drawn from a binomial distribution, and satellite galaxies drawn from a Poisson distribution conditioned to the presence of a central galaxy
(we will not consider the issue of sample selection here).
More specifically, one writes the average number of galaxies in a halo as 
\be 
\langle N_g \rangle = \langle N_\mr{cen} \rangle + \langle N_\mr{sat} \rangle
\ee
and we will adopt \cite{Tinker:2009mx, Penin:2013zya,Lacasa2013}
\be 
\langle N_\mr{cen} \rangle = \frac{1}{2} \left[ 1 + \mbox{erf} \left( \frac{\log M - \log M_{min}}{\sigma_{\log M}} \right) \right]
\ee
and
\be  
\langle N_\mr{sat} \rangle = \frac{1}{2} \left[ 1 + \mbox{erf} \left( \frac{\log M - \log 2M_{min}}{\sigma_{\log M}} \right) \right] \left(\frac{M}{M_\mr{sat}}\right)^{\alpha_\mr{sat}}
\ee
This parametrization has 4 parameters:
a mass threshold above which a halo has a large probability of containing a central galaxy ($M_{min}$),
the width of the transition of the central probability ($\sigma_{\log M}$),
the typical mass above which a halo contains satellite galaxies ($M_\mr{sat}$) and the index of the power law for the number of satellites at large halo masses ($\alpha_\mr{sat}$).

Finally, we use the following mathematical notations. 
Estimators of a given quantity are noted with an overhat, and overbar denote the corresponding average. 
We also use the comoving volume per solid angle, defined as
\be 
\dd V = r^2(z) \, \frac{\dd r}{\dd z} \, \dd z ,
\ee
as integration element for simplicity, with $r(z)$ being the comoving distance.\\
We may also denote only with indices the multiplication of some quantities, e.g.:
\be
\dd V_{12} = \dd V_1 \ \dd V_2 = r^2(z_1) \left.\frac{\dd r}{\dd z}\right|_{z_1} \ r^2(z_2) \left.\frac{\dd r}{\dd z}\right|_{z_2} \dd z_1 \,\dd z_2.
\ee

Otherwise stated, we will adopt a flat $\Lambda$CDM model with the cosmological parameters $\Omega_b h^2 = 0.022$, $\Omega_{cdm} h^2 = 0.12$,
$h=0.67$, $n_s = 0.962$, $\sigma_8 = 0.834$ and use the parametrization of Eisenstein and Hu \cite{Eisenstein:1997ik} to generate the linear matter power spectrum.
For HOD parameters we use $\alpha_\mr{sat} = 1.3$, $M_\mr{min} = 10^{12.2} \Msun$ and $M_\mr{sat} = 10^{13.2} \Msun$ and $\sigma_{\log M} = 0.2$~\cite{Penin:2013zya}.

\section{Cluster number counts and its covariance}\label{Sect:counts}

Cluster number counts is an important cosmological observable that is sensitive to both
the  background evolution as well as the growth of perturbations. Cluster counts are also
independent of the cluster density profile and HOD modelling, depending exclusively on the cluster mass function.

The estimator of the binned cluster number counts per unit of solid angle 
is simply the number of clusters in a given mass and redshift bin divided by the survey area, which we take to be the full sky from now,  $\Omega_S = 4\pi$. It can be written as:
\be\label{Eq:counts_binned_estimator_sumj}
\hat{N}_\mr{cl}(i_M,i_z) = \frac{1}{4\pi} \int \dd^2\hn \, \dd M \, \dd z \ \sum_j \delta(M-M_j) \, \delta^2(\hn-\hn_j) \, \delta(z-z_j),
\ee
where the mass and redshift integrals run over the bins defined by $i_M$ and $i_z$ and the index $j$ runs over all halos in the universe with mass $M_j$ and centered at
$\vec{x}_j = r(z_j) \hat{n}_j$.
We assume that all clusters can be detected without fake detections and that both their mass and redshift can be measured perfectly. The effects of the spread in the mass-observable relation for the cluster mass, the purity and completeness of the cluster catalogue and of the photometric redshift errors will be considered in future works.

\begin{figure}[htb]
\centering
\includegraphics[width=.8\linewidth]{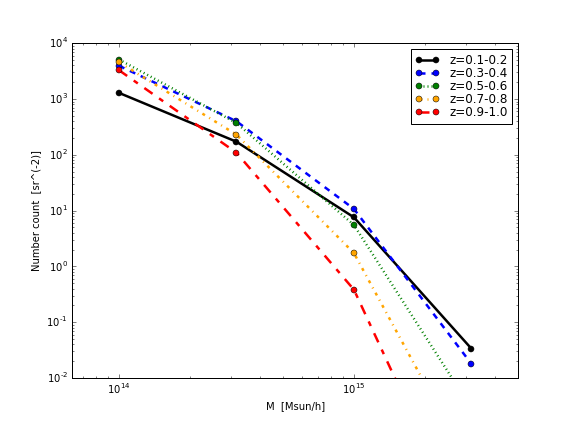}
\caption{Cluster number counts in five representative redshift bins. The mass bins have width $\Delta\log M=0.5$ and the points are drawn at the start of the mass bin.}
\label{Fig:Number-count}
\end{figure}

Eq.~(\ref{Eq:counts_binned_estimator_sumj}) can be rewritten as:
\ba \label{Eq:counts_binned_estimator_compact}
\hat{N}_\mr{cl}(i_M,i_z) &= \frac{1}{4\pi} \int \dd^2\hn \, \dd M \, \dd V \,  {\cal N}_{cl}(\xx=r\hn | M,z)
\ea
and hence the average of the cluster number count is given by
\be
\bar{N}_\mr{cl}(i_M,i_z) = \lbra \hat{N}_\mr{cl}(i_M,i_z) \rbra = \int \dd M \, \dd V \, \frac{\dd n_\mr{h}}{\dd M} 
\ee
where the mass and redshift integrals implicitly run over the corresponding bins. The resulting cluster counts are shown in figure \ref{Fig:Number-count}.

Eq. (\ref{Eq:counts_binned_estimator_compact}) can be rewritten in harmonic space as
\ba
\hat{N}_\mr{cl}(i_M,i_z) &=  \frac{1}{4\pi} \int \frac{\dd^3\kk}{(2\pi)^3} \, \dd^2\hn \, \dd M \, \dd V \ {\cal N}_{cl}(\kk | M,z) \, e^{i\kk\cdot\xx} \\
\label{Eq:counts_binned_estimator_k} &= \int \frac{\dd^3\kk}{(2\pi)^3} \, \dd M \, \dd V \ j_0(kr) \, {\cal N}_{cl}(\kk | M,z)
\ea
where we used the plane wave expansion $e^{i\kk\cdot\xx} = 4\pi \sum_{\ell m} i^\ell j_\ell(kr) \, Y^*_{\ell m}(\hk) \, Y_{\ell m}(\hn)\,$, and the integral over the sphere leaves only the monopole ($\ell=0$) contribution.

From eq.~(\ref{Eq:counts_binned_estimator_k}), the covariance of cluster number counts is given by
\be
\nonumber \mr{Cov}\left(\hat{N}_\mr{cl}(i_M,i_z),\hat{N}_\mr{cl}(j_M,j_z)\right) = \int \frac{\dd^3\kk}{(2\pi)^3} \, \dd M_{12} \, \dd V_{12} \ j_0(k r_1) \, j_0(k r_2) 
\left.\frac{\dd n_h}{\dd M}\right|_{M_1,z_1} \left.\frac{\dd n_h}{\dd M }\right|_{M_2,z_2} \ P_\mr{cl}(k | M_{12}, z_{12})
\ee
where the cluster power spectrum 
\be 
P_\mr{cl}(k | M_{12}, z_{12}) = (2 \pi)^3 \delta^3(\kk_1 - \kk_2) \langle \delta_{h}(\kk_1 | M_1,z_1) \delta_{h}(\kk_2 | M_2,z_2) \rangle
\ee
has contributions from the 1- and 2-halo terms and
we have introduced the halo density contrast $\delta_h$ such that 
\begin{equation}
\mathcal{N}_\mr{cl}(\kk | M,z)=\left.\frac{\dd n_h}{\dd M}\right|_{M,z} \delta_h(\kk|M,z). 
\label{Eq:deltah}
\end{equation}

The 1-halo term contribution is diagonal in the mass and redshift bins and is inversely proportional to the survey area. It's contribution to the covariance is given by:
\be
\nonumber \mr{Cov}_\mr{1h}\left(\hat{N}_\mr{cl}(i_M,i_z),\hat{N}_\mr{cl}(j_M,j_z)\right)  =
\delta_{i_M,j_M} \, \delta_{i_z,j_z} \ \frac{\bar{N}_\mr{cl}(i_M,i_z)}{4\pi},
\ee
and it represents a Poissonian shot-noise error. 

The 2-halo contribution to the number counts covariance can be written as:
\ba
\nonumber \mr{Cov}_\mr{2h}\left(\hat{N}_\mr{cl}(i_M,i_z),\hat{N}_\mr{cl}(j_M,j_z)\right) &= \int \dd M_{12} \, \dd V_{12} \, \left.\frac{\dd n_h}{\dd M}\right|_{M_1,z_1} \left.\frac{\dd n_h}{\dd M }\right|_{M_2,z_2} b_1(M_1,z_1) \, b_1(M_2,z_2) \\
& \qquad \times \int \frac{\dd^3\kk}{(2\pi)^3} \; j_0(k r_1) \, j_0(k r_2) \; P_\mr{lin}(k | z_{12})
\ea
where $ P_\mr{lin}(k | z_{12})$ is the linear matter power spectrum and we have assumed that halos are linearly biased with respect to matter. \\

This equation can be rewritten as:
\be
\label{Eq:Cov-counts-2h-SSCform} \mr{Cov}_\mr{2h}\left(\hat{N}_\mr{cl}(i_M,i_z),\hat{N}_\mr{cl}(j_M,j_z)\right) = \int \dd V_{12} \ \frac{\partial n_h}{\partial\delta_b}(i_M,z_1) \;   \frac{\partial n_h}{\partial\delta_b} (j_M,z_2) \ \sigma_\mr{proj}^2(z_1,z_2)
\ee
with
\be
\frac{\partial n_h}{\partial\delta_b}(i_M,z)  \equiv \int_{M \in \mr{bin}(i_M)} \dd M \ \frac{\dd n_h}{\dd M } \ b_1(M,z),
\label{Eq:deltan}
\ee
and
\ba 
\sigma_\mr{proj}^2(z_1,z_2) \equiv \int \frac{\dd^3\kk}{(2\pi)^3} \; j_0(k r_1) \, j_0(k r_2) \; P_\mr{lin}(k | z_{12}).
\ea
The significance of this form of writing the result and its relation to sample variance will become clear in the discussion presented in section \ref{Sect:SSC}.

\begin{figure}[htb]
\centering
\includegraphics[width=.7\linewidth]{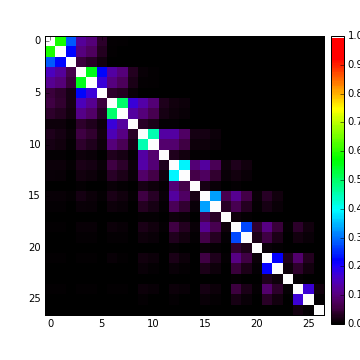}
\caption{Cluster number counts covariance for 3 mass bins and 9 redshift bins. See text for details.}
\label{Fig:Cov-Ncl}
\end{figure}

This 2-halo contribution is off-diagonal: it couples different mass bins and different redshift bins.
Figure \ref{Fig:Cov-Ncl} shows the normalised cluster counts covariance, in absolute value, obtained with the following set up: three mass bins of width $\Delta \log M = 0.5$ between $10^{14} \Msun$ and $10^{15.5} \Msun$ and nine redshift bins of width $\Delta z=0.1$ between $z=0.1$ and $1$.
We plotted the absolute value of the covariance for clarity, as the cross-covariance between neighbouring redshift bins is negative due to the behaviour of 
$\sigma_\mr{proj}^2(z_1,z_2)$ (shown in figure \ref{Fig:sigmaproj} of section \ref{Sect:SSC}). 
The covariance shows significant correlation between the two lowest mass bins, and the off-diagonality increases at lower redshift, to the left, as expected from the larger nonlinear effects.

\section{Galaxy angular power spectrum}\label{Sect:galspec}

The projected galaxy density contrast\footnote{Note that we normalise the contrast by the theoretical average $\overline{n}_\mr{gal}$, instead of the often used empirical average over the survey. Indeed we note that normalising by the empirical average yields an ill-defined statistical quantity, as a ratio of estimators, which is in particular biased. Normalising by the theoretical average from the model is perfectly feasible too and yields constraints identical to those using the empirical absolute difference $n_\mr{gal}-\overline{n}_\mr{gal}$. We leave for future works the question of dealing with all the problems of the empirical normalisation.} 
in the direction $\hn$ in a redshift bin $i_z$, $\delta_\mr{gal}(\hn,i_z)$ is:
\be\label{Eq:dngalbinned_rig}
\delta_\mr{gal}(\hn,i_z) 
= \int_{i_z} \dd V \; \frac{\overline{n}_\mr{gal}(z) }{\Delta N_\mr{gal}(i_z)} \ \delta_\mr{gal}(\xx=r\hn,z)
\ee
where we denote as $\overline{n}_\mr{gal}(z)$ the 3D average galaxy density at redshift $z$ modelled by
\be 
\overline{n}_\mr{gal}(z) = 
\int_{i_z} \dd V \dd M\; \frac{d n_h}{dM} \langle N_g \rangle,
\ee
and where the number of galaxies per steradian in the redshift bin $i_z$ is given by
\be 
\Delta N_\mr{gal}(i_z) = \int_{i_z} dV \ \overline{n}_\mr{gal}(z).
\ee

After spherical harmonic decomposition, the harmonic coefficients are:
\ba 
a_{\ell m}^\mr{gal}(i_z)  &= \int \dd^2\hn \; \delta_\mr{gal}(\hn,i_z)\; Y^*_{\ell m}(\hn)\\
&= \int \dd V \; \frac{\overline{n}_\mr{gal}(z) }{\Delta N_\mr{gal}(i_z)} \int \dd^2\hn \, \frac{\dd^3\kk}{(2\pi)^3} \; \delta_\mr{gal}(\kk,z) \; e^{i \kk \cdot r\hn} \; Y^*_{\ell m}(\hn)\\
&= i^\ell \int \dd V \; \frac{\overline{n}_\mr{gal}(z) }{\Delta N_\mr{gal}(i_z)} \int \frac{\dd^3\kk}{2\pi^2} \; j_\ell(k r) \; \delta_\mr{gal}(\kk,z) \; Y^*_{\ell m}(\hk)
\ea
The galaxy power spectrum estimator is then:
\ba
\hat{C}_\ell^\mr{gal}(i_z,j_z) &= \frac{1}{2\ell+1} \sum_m \ a_{\ell m}^\mr{gal}(i_z) \ \left(a_{\ell m}^\mr{gal}(j_z)\right)^* \\
\label{Eq:Clgal-estimator-Lagrange} \nonumber &= 4\pi \int \dd V_{12} \; \frac{\overline{n}_\mr{gal}(z_1) \, \overline{n}_\mr{gal}(z_2)}{\Delta N_\mr{gal}(i_z) \, \Delta N_\mr{gal}(j_z)} \int \frac{\dd^3\kk_{12}}{(2\pi)^6} \; j_\ell(k_1 r_1) \,  j_\ell(k_2 r_2)\\
& \qquad \times P_\ell(\hk_1 \cdot \hk_2) \ \delta_\mr{gal}(\kk_1,z_1) \; \delta^*_\mr{gal}(\kk_2,z_2)
\ea
It's average is
\ba
\nonumber \bar{C}_\ell^\mr{gal}(i_z,j_z) &= 4\pi \int \dd V_{12} \; \frac{\overline{n}_\mr{gal}(z_1) \, \overline{n}_\mr{gal}(z_2)}{\Delta N_\mr{gal}(i_z) \, \Delta N_\mr{gal}(j_z)} \int \frac{\dd^3\kk}{(2\pi)^3} \; j_\ell(k r_1) \,  j_\ell(k r_2) \; P_\mr{gal}(k | z_{12})\\
&= \frac{2}{\pi} \int \dd V_{12} \; \frac{\overline{n}_\mr{gal}(z_1) \, \overline{n}_\mr{gal}(z_2)}{\Delta N_\mr{gal}(i_z) \, \Delta N_\mr{gal}(j_z)} \int k^2 \, \dd k \; j_\ell(k r_1) \,  j_\ell(k r_2) \; P_\mr{gal}(k | z_{12}) \\
& \simeq 
\frac{\delta_{i_z,j_z}}{\Delta N_\mr{gal}(i_z)^2} \int \dd V \ \overline{n}_\mr{gal}(z)^2 \, P_\mr{gal}(k_\ell | z)
\ea
where in the last step we have used 
Limber's approximation, taking the power spectrum  at the maximum of the Bessel functions, $k_\ell = \frac{\ell+1/2}{r(z)}$, and
performing the integration of the Bessel functions that yields a Dirac delta function in redshift 
(for more details on the Limber approximation for polyspectra at any order, see appendix E of \cite{Lacasa2014a}).
The 3-d galaxy power spectrum is defined in the usual manner
\be 
P_\mr{gal}(k| z_{12}) = 
(2 \pi)^3 \delta^3 (\kk - \kk') \langle \delta_\mr{gal}(\kk,z_1) \; \delta^*_\mr{gal}(\kk',z_2) \rangle.
\ee

We can use the Halo Model to write the galaxy power spectrum as a sum of a 2-halo, a 1-halo and a shot-noise term:
\be
P_\mr{gal}(k | z_{12}) = P^\mr{2h}_\mr{gal}(k | z_{12}) + P^\mr{1h}_\mr{gal}(k | z_{12}) + P^\mr{shot}_\mr{gal}(k | z_{12})
\ee
with
\ba
P^\mr{2h}_\mr{gal}(k | z_{12}) &= b_1^\mr{gal,eff}(k,z_1) \, b_1^\mr{gal,eff}(k,z_2) \, P_\mr{DM}(k | z_{12}) \\
P^\mr{1h}_\mr{gal}(k | z_{12}) &= \frac{\delta_{z_1,z_2}}{\overline{n}_\mr{gal}(z_1)^2} \int \dd M \, \frac{\dd n_h}{\dd M} \, \lbra N_\mr{gal} (N_\mr{gal}-1)(M)\rbra\, u(k | M,z_1)^2 \\
P^\mr{shot}_\mr{gal}(k | z_{12}) &= \frac{\delta_{z_1,z_2}}{\overline{n}_\mr{gal}(z_1)}
\ea
where $b^\mr{gal,eff}_1$ is the (first-order) effective galaxy bias:
\be
b^\mr{gal,eff}_1(k, z) = \int \dd M \left.\frac{\dd n_\mr{h}}{ \dd M}\right|_{M,z} \!\!\!\! \frac{\lbra N_\mr{gal}(M) \rbra}{\overline{n}_\mr{gal}(z)} u(k | M) \, b_1(M,z).
\ee
Note that on large scales, $u(k | M) \rightarrow 1$ and the effective bias goes to a constant, the usual galaxy bias: 
$b^\mr{gal}_1(z) = \int \dd M \, \frac{\dd n_\mr{h}}{\dd M} \, \frac{\lbra N_\mr{gal}(M) \rbra}{\overline{n}_\mr{gal}(z)} \; b_1(M,z)$.

\begin{figure}[htb]
\centering
\includegraphics[width=1.\linewidth]{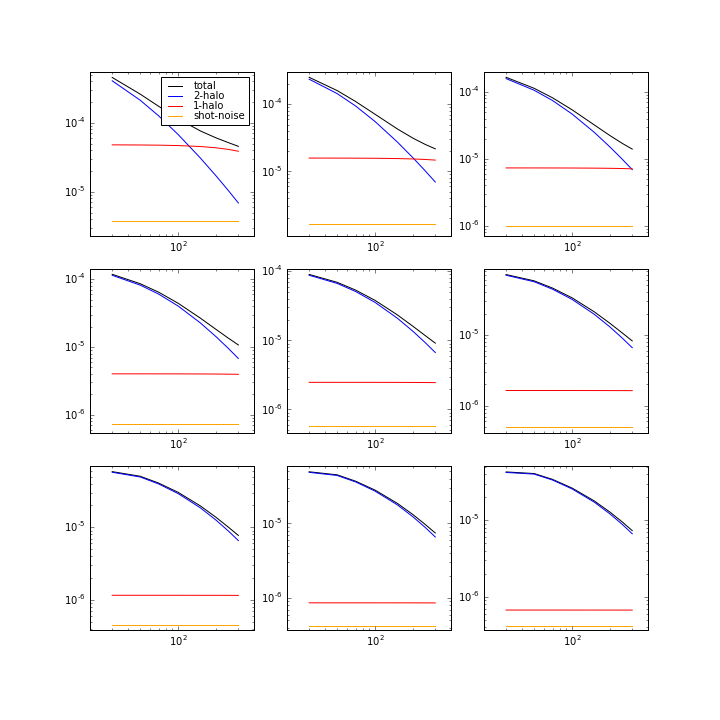}
\caption{Galaxy angular power spectrum and its different terms on large scales $\ell=30-300$. Each subplot represent a redshift bin of width $\Delta z=0.1$, starting at $z=0.1-0.2$ at the upper left and ending at $z=0.9-1$ at the bottom right.}
\label{Fig:Clgal}
\end{figure}

The resulting galaxy power spectrum and its different terms are shown in figure \ref{Fig:Clgal}.
The shot-noise term reaches at best 8\% of the total spectrum on these scales. 
On the other hand, the 1-halo term becomes fairly important at low redshift, dominating at the (relatively) small scales, as expected.

\section{Cross-covariance between cluster counts and galaxy power spectrum}\label{Sect:crosscov}

We now turn to the modelling of the cross-covariance between the number counts studied in section \ref{Sect:counts} and the galaxy spectrum studied in section \ref{Sect:galspec}. 
It is indeed the main goal of this study to estimate the impact of including this cross-covariance in the extraction of cosmological and astrophysical parameters from large galaxy surveys.

Following eq.~(\ref{Eq:counts_binned_estimator_k})  and eq.~(\ref{Eq:Clgal-estimator-Lagrange}), the covariance between the number counts and the galaxy power spectrum estimators is:
\ba
\nonumber \Cov\left(\hat{N}_\mr{cl}(i_M,i_z) , \hat{C}_\ell^\mr{gal}(j_z,k_z)\right) &= 4\pi \int \frac{\dd^3\kk_{123}}{(2\pi)^9} \, \dd M_1 \, \dd V_{123} \; \frac{\nbargal(z_2) \, \nbargal(z_3)}{\Delta N_\mr{gal}(j_z)^2} \; P_\ell(\hk_2\cdot\hk_3) \\
\label{Eq:cov-Ncl-Clgal-estim} & \times j_0(k_1 r_1) \, j_\ell(k_2 r_2) \, j_\ell(k_3 r_3) \; \lbra \mathcal{N}_\mr{cl}(\kk_1 | M_1,z_1) \, \delta_\mr{gal}(\kk_2,z_2) \, \delta^*_\mr{gal}(\kk_3,z_3) \rbra_c.
\ea

Therefore the cross covariance is related to a 3-point function involving one halo/cluster and two galaxies, the so-called bispectrum defined by:
\be 
\langle \delta_h(\kk_1|M_1,z_1) \, \delta_g(\kk_2,z_2) \, \delta_g(\kk_3,z_3) \rangle_c = (2 \pi)^3 \, \delta^3(\kk_1+\kk_2+\kk_3) \, B_\mr{hgg}(k_{123}| M_1, z_{123}),
\ee
where we used the halo density contrast $\delta_h$ defined in eq.~(\ref{Eq:deltah}).

\begin{figure}[htb]
\centering
\includegraphics[width=.8\linewidth]{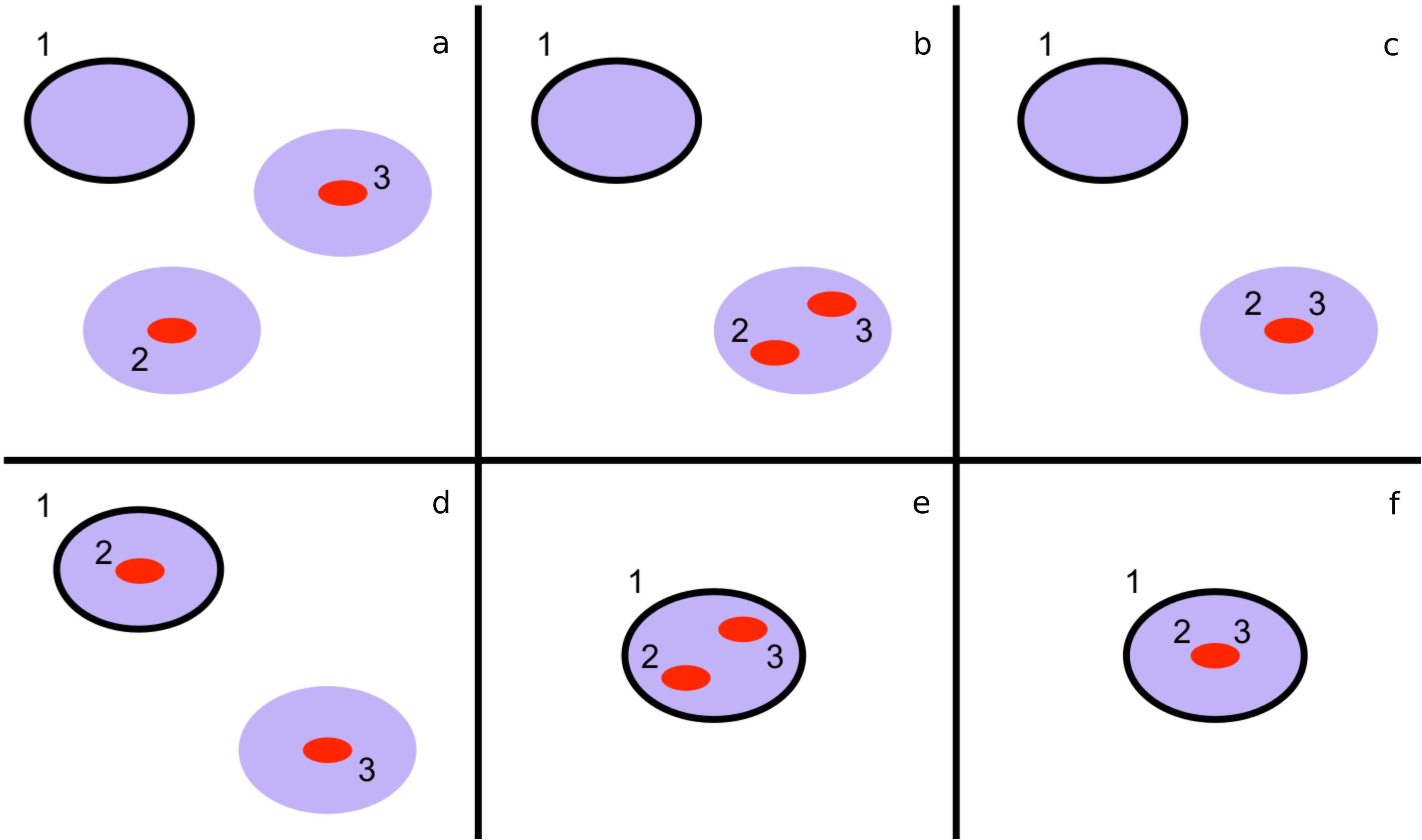}
\caption{Diagrams for the halo-galaxy-galaxy bispectrum. From left to right and top to bottom : 3h, 2h-1h2g, 2h-1h1g, 2h-2h, 1h2g, 1h1g.}
\label{Fig:diagrams}
\end{figure}

As done for the galaxy power spectrum, we model this halo-galaxy-galaxy bispectrum using the halo model coupled with a Halo Occupation Distribution. 
There are six terms involved and they can be easily visualised with simple diagrams, as first proposed by Lacasa et al.~\cite{Lacasa2013}, which are shown in figure~\ref{Fig:diagrams}.

The first three diagrams (upper row) can couple number counts and galaxy spectrum in different redshift bins, and we will later identify with super-sample covariance.
They are described as:
\begin{enumerate}[a)]
\setlength{\itemsep}{0pt}
\item the 3h term (upper left diagram) quantifies how halo counts and the 2-halo part of the galaxy spectrum fluctuate together with large scale structure. It involves the halo bispectrum, which has three contributions : the dark matter bispectrum from second order perturbation theory (2PT term), second order halo bias (b2 term), and non-local halo bias (s2 term). 
\item the 2h-1h2g term (upper central diagram) quantifies how halo counts and the 1-halo part of the galaxy spectrum fluctuate together with large scale structure.
\item the 2h-1h1g term (upper right diagram) quantifies how halo counts and the shot-noise part of the galaxy spectrum fluctuate together with large scale structure.
\end{enumerate}

The three diagrams in the bottom row contribute only when the number count and the galaxy spectrum are in the same redshift bin and are described as:
\begin{enumerate}[a)]
\setlength{\itemsep}{0pt}
\setcounter{enumi}{3}
\item the 2h-2h term (bottom left diagram) quantifies how the halos in that bin of mass source the 2-halo part of the galaxy power spectrum.
\item the 1h2g term (bottom central diagram) quantifies how the halos in that bin of mass source the 1-halo part of the galaxy power spectrum.
\item the 1h1g term (bottom right diagram) quantifies how the halos in that bin of mass source the shot-noise part of the galaxy power spectrum.
\end{enumerate}

Therefore we can write:
\be
B_\mr{hgg}(k_{123},z_{123} | M_1,z_{123}) = B^{3h} + B^{2h-2h} + B^{2h-1h2g} + B^{2h-1h1g} + B^{1h2g} + B^{1h1g}
\ee
with
\be 
B^{3h} = B^{2PT} + B^{b2} + B^{s2}. 
\ee 
The expressions for each of these terms are given in appendix \ref{App:3Dhgg}.

With these notations, we show in appendix \ref{App:crosscov-deriv} that the cross covariance reduces in the Limber approximation to
\ba
\nonumber \mr{Cov}\left(\hat{N}_\mr{cl}(i_M,i_z),C_\ell(j_z,k_z)\right) &= \frac{\delta_{j_z,k_z}}{4\pi} \int \frac{\dd M_1 \, \dd V_{12}}{\Delta N_\mr{gal}(j_z)^2} \left.\frac{\dd n_h}{\dd M}\right|_{M_1,z_1} \overline{n}_\mr{gal}(z_2)^2 \\
\label{Eq:crosscov-wlimber} & \times \frac{2}{\pi} \int k^2_1 \dd k_1 \, B_\mr{hgg}(k_1,k_\ell,k_\ell | M_1,z_{122}) \, j_0(k_1 r_1) \, j_0(k_1 r_2)
\ea
and the different cross covariance terms are:
\ba
\label{Eq:CrossCov-1h1g} \mr{Cov}_\mr{1h1g}\left(\hat{N}_\mr{cl}(i_M,i_z),\hat{C}_\ell^\mr{gal}(j_z)\right) &\approx \frac{\delta_{i_z,j_z}}{4\pi} \int \dd V_1 \ \frac{n_\mr{gal}(z_1 | i_M)}{\Delta N_\mr{gal}(i_z)^2} \\
\label{Eq:CrossCov-1h2g} \mr{Cov}_\mr{1h2g}\left(\hat{N}_\mr{cl}(i_M,i_z),\hat{C}_\ell^\mr{gal}(j_z)\right) &= \frac{\delta_{i_z,j_z}}{4\pi} \int \dd V_1 \ \frac{\overline{n}_\mr{gal}(z_1)^2 \; P_\mr{gal}^\mr{1h}(k_\ell , z_1 | i_M)}{\Delta N_\mr{gal}(i_z)^2} \\
\nonumber \mr{Cov}_\mr{2h-2h}\left(\hat{N}_\mr{cl}(i_M,i_z),\hat{C}_\ell^\mr{gal}(j_z)\right) &= \frac{\delta_{i_z,j_z}}{4\pi} \int \dd V_1 \ 2 \; b^\mr{gal,eff}_1(k_\ell,z_1 | i_M) \; b^\mr{gal,eff}_1(k_\ell,z_1) \\
\label{Eq:CrossCov-2h-2h} & \qquad \qquad \times \frac{\overline{n}_\mr{gal}(z_1)^2}{\Delta N_\mr{gal}(i_z)^2} \; P_\mr{DM}(k_\ell,z_1)\\
\nonumber \mr{Cov}_\mr{2h-1h1g}\left(\hat{N}_\mr{cl}(i_M,i_z),\hat{C}_\ell^\mr{gal}(j_z)\right) &= \int \frac{\dd V_{12} }{\Delta N_\mr{gal}(j_z)^2} \; \frac{\partial n_h}{\partial \delta_b}(i_M,z_1) \\
\label{Eq:CrossCov-2h-1h1g} & \qquad \qquad \times \overline{n}_\mr{gal}(z_2) \, b^\mr{gal,eff}_1(z_2) \; \sigma^2_\mr{proj}(z_1,z_2) \\
\nonumber \mr{Cov}_\mr{2h-1h2g}\left(\hat{N}_\mr{cl}(i_M,i_z),\hat{C}_\ell^\mr{gal}(j_z)\right) &= \int \frac{\dd V_{12}}{\Delta N_\mr{gal}(j_z)^2} \; \frac{\partial n_h}{\partial \delta_b}(i_M,z_1) \\
\label{Eq:CrossCov-2h-1h2g} & \qquad \qquad \times \overline{n}_\mr{gal}(z_2) \, b_1^\mr{gal,eff-2}(k_\ell , z_2) \; \sigma^2_\mr{proj}(z_1,z_2) \\
\label{Eq:CrossCov-s2} \mr{Cov}_\mr{s2}\left(\hat{N}_\mr{cl}(i_M,i_z),\hat{C}_\ell^\mr{gal}(j_z)\right) &= 0 \\
\nonumber \mr{Cov}_\mr{b2}\left(\hat{N}_\mr{cl}(i_M,i_z),\hat{C}_\ell^\mr{gal}(j_z)\right) &=
\int \frac{\dd V_{12}}{\Delta N_\mr{gal}(j_z)^2} \, 2 \, \overline{n}_\mr{gal}(z_2)^2 \, b_1^\mr{gal,eff}(k_\ell ,z_2) \, b_2^\mr{gal,eff}(k_\ell ,z_2) \\
\label{Eq:CrossCov-b2} & \quad \times \frac{\partial n_h}{\partial \delta_b}(i_M,z_1) \, P_\mr{DM}(k_\ell |z_2) \; \sigma^2_\mr{proj}(z_1,z_2) \\
\nonumber \mr{Cov}_\mr{2PT}\left(\hat{N}_\mr{cl}(i_M,i_z),\hat{C}_\ell^\mr{gal}(j_z)\right) &= \int \frac{\dd V_{12}}{\Delta N_\mr{gal}(j_z)^2} \, 4 \, F_\mr{avg} \; \overline{n}_\mr{gal}(z_2)^2 \, b_1^\mr{gal,eff}(k_\ell ,z_2)^2 \\
\label{Eq:CrossCov-2PT} & \quad \times \frac{\partial n_h}{\partial \delta_b}(i_M,z_1) \, P_\mr{DM}(k_\ell |z_2) \; \sigma^2_\mr{proj}(z_1,z_2),
\ea
where we re-used notations of the previous sections, and introduced the following additional short-hand notations (notice that some of the integrals are on a given mass bin) :
\ba
n_\mr{gal}(z | i_M) &= \int_{M\in\mr{bin}(i_M)} \!\!\!\!\!\!\dd M \ \frac{\dd n_h}{\dd M} \ \lbra N_\mr{gal}(M)\rbra \\
P^\mr{1h}_\mr{gal}(k, z | i_M) &= \frac{1}{\nbargal^2(z)} \int_{M\in\mr{bin}(i_M)} \!\!\!\!\!\!\dd M \ \frac{\dd n_h}{\dd M} \ \lbra N_\mr{gal}(N_\mr{gal}-1)(M)\rbra \ u^2(k|M,z)\\
b_{1,2}^\mr{gal,eff}(k, z | i_M) &= \frac{1}{\nbargal(z)} \int_{M\in\mr{bin}(i_M)} \!\!\!\!\!\!\dd M \ \frac{\dd n_h}{\dd M} \ \lbra N_\mr{gal}(M)\rbra \ b_{1,2}(M,z) \ u(k|M,z)\\
b_1^\mr{gal,eff-2}(k, z) &= \frac{1}{\nbargal(z)} \int \dd M \ \frac{\dd n_\mr{h}}{\dd M} \ \lbra N_\mr{gal}(N_\mr{gal}-1)(M)\rbra \ b_1(M,z) \ u(k|M,z)^2\\
F_\mr{avg} &= \frac{17}{21}
\ea

\begin{figure}[htb]
\centering
\includegraphics[width=.55\linewidth]{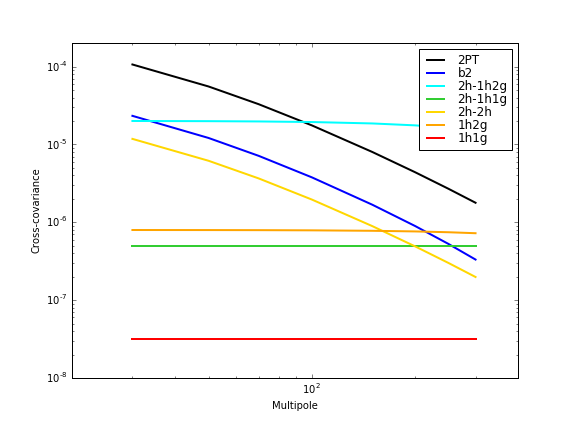}\\
\includegraphics[width=.55\linewidth]{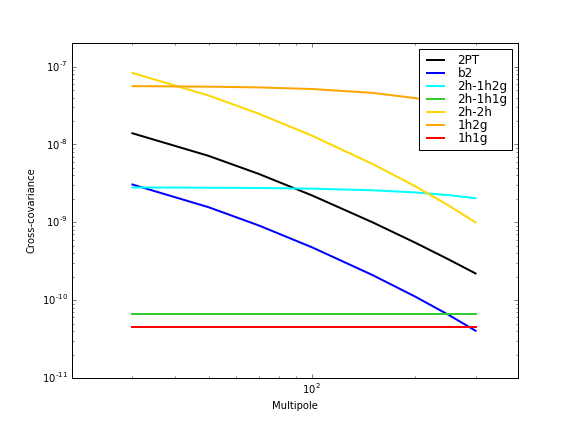}\\
\includegraphics[width=.55\linewidth]{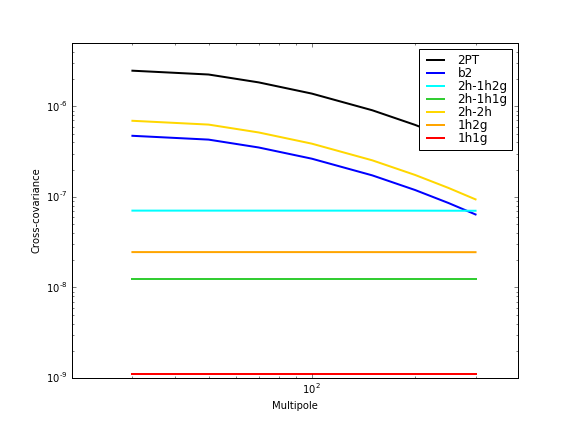}
\caption{Cross-covariance between cluster counts and the galaxy spectrum, depending on the multipole of the spectrum. \textit{Top:} low redshift low mass case, $z\in [0.1,0.2]$ $\log M \in [14,14.5]$. \textit{Center:} low redshift high mass case, $z\in [0.1,0.2]$ $\log M \in [15,15.5]$. \textit{Bottom:} high redshift low mass case, $z\in [0.8,0.9]$ $\log M \in [14,14.5]$.}
\label{Fig:CrossCov}
\end{figure}

The resulting cross-covariance is shown in figure \ref{Fig:CrossCov} as a function of multipole $\ell$, with three plots for different cases of mass bin and redshift bin. In these plots we take $i_z=j_z$ for simplicity and because this is where the cross-covariance is maximal.

We see from figure \ref{Fig:CrossCov} that different terms become important depending on scale, mass and redshift. In particular HOD terms are far from negligible, being dominant in some cases, while the 2PT term can become negligible in some cases. We note that the 2PT term is the only term that one would have considered from a model with only linear bias and perturbation theory. 
We thus conclude that it is critical to use a fully non-linear model, such as the halo model combined with HOD that we have been using here, in order to estimate accurately the cross-covariance.

\section{Super-sample covariance}\label{Sect:SSC}

We can unify the equations (\ref{Eq:CrossCov-2h-1h1g},\ref{Eq:CrossCov-2h-1h2g},\ref{Eq:CrossCov-b2},\ref{Eq:CrossCov-2PT}) for the 2h-1h1g, 2h-1h2g, b2 and 2PT 
terms involvong the super-sample covariance, into a single equation:
\ba
\mr{Cov}_\mr{SSC}\left(\hat{N}_\mr{cl}(i_M,i_z),\hat{C}_\ell^\mr{gal}(j_z)\right) &\equiv \mr{Cov}_\mr{2h-1h1g} + \mr{Cov}_\mr{2h-1h2g} + \mr{Cov}_\mr{b2} + \mr{Cov}_\mr{2PT}\\
\label{Eq:CrossCov-SSC} & = \int \dd V_{12} \, \frac{\nbargal(z_2)^2}{\Delta N_\mr{gal}(j_z)^2} \, \frac{\partial n_h}{\partial\delta_b}(i_M,z_1) \, \frac{\partial P_\mr{gal}(k_\ell | z_2)}{\partial \delta_b} \, \sigma^2_\mr{proj}(z_1,z_2)
\ea
if we define
\ba
\nonumber \frac{\partial P_\mr{gal}(k | z)}{\partial \delta_b} &\equiv \underbrace{4 F_\mr{avg}}_{=\frac{68}{21}} \, b_1^\mr{gal,eff}(k,z)^2 \, P_\mr{DM}(k|z)+ 
b_1^\mr{gal,eff-2}(k,z)/\nbargal(z)  \\
\label{Eq:dPgalddeltab} &+ 2 b_1^\mr{gal,eff}(k,z) \, b_2^\mr{gal,eff}(k,z) \, P_\mr{DM}(k|z) + b_1^\mr{gal,eff}(z)/\nbargal(z).
\ea
These terms correspond to the diagrams in the top row of figure \ref{Fig:diagrams} and represent 
how cluster counts and the galaxy power spectrum fluctuate together due to the large scale structure.

By analogy with previous literature results for the 3D matter power spectrum \cite{Takada:2013wfa}, eq.~(\ref{Eq:dPgalddeltab}) can be interpreted as the reaction of the 
galaxy power spectrum to a change of background density and hence the notation $\frac{\partial P_\mr{gal}}{\partial \delta_b}$. 
We also interpret $\frac{\partial n_h}{\partial\delta_b}$, defined in eq.~(\ref{Eq:deltan}) as the reaction of the 
cluster number counts to a change of background density.

The first two terms in eq.~(\ref{Eq:dPgalddeltab}) correspond to the expression found in \cite{Takada:2013wfa}. However, in contrast to \cite{Takada:2013wfa} 
we have here derived our equations without any Taylor expansion nor approximation, except for Limber's approximation. 
Furthermore, we have derived here the contributions from second order bias (third term), galaxy shot-noise (fourth term), 
and we have shown that non-local tidal effects give a zero contribution (see appendix \ref{App:crosscov-deriv}  for details).
In our framework, eq.~(\ref{Eq:CrossCov-SSC}) can be interpreted physically as the fact that both cluster counts and the galaxy spectrum 
react to a change of background, so their covariance is related to the covariance of the background density $\sigma^2_\mr{proj}(z_1,z_2)$. Indeed:
\ba
\sigma^2_\mr{proj}(z_1,z_2) &= \int \frac{\dd^3\kk}{(2\pi)^3} \; j_0(k r_1) \, j_0(k r_2) \, P_\mr{DM}(k | z_{12}) \\
&= \lbra \delta_\mr{avg}(z_1) \, \delta_\mr{avg}(z_2)\rbra_c \quad \mr{with} \quad \delta_\mr{avg}(z) = \int \frac{\dd^2\hn}{4\pi} \, \delta(r\hn,z).
\ea

This source of covariance has been dubbed ``super-sample covariance" in the literature, as it gives a non-diagonal covariance due to modes larger than the survey. 
In our case the survey is the whole sphere and the Bessel function $j_0$ indicates that the contributing Fourier modes are those associated with the monopole.
Eq.~(\ref{Eq:Cov-counts-2h-SSCform}) for the 2-halo term of the covariance of cluster counts can also be interpreted as a super-sample covariance contribution, 
though it is more often more simply called ``sample variance" in the cluster literature.

As a generalisation, it is natural to expect that the covariance of two observables $\mathcal{O}_1$ and $\mathcal{O}_2$ will have a contribution of the form:
\be
\Cov_\mr{SSC}\left(\mathcal{O}_1,\mathcal{O}_2\right) = \int \dd V_{12} \, \frac{\partial \mathcal{O}_1}{\partial\delta_b}(z_1) \, \frac{\partial \mathcal{O}_2}{\partial \delta_b}(z_2) \, \sigma^2_\mr{proj}(z_1,z_2)
\ee

\begin{figure}[htb]
\centering
\includegraphics[width=.7\linewidth]{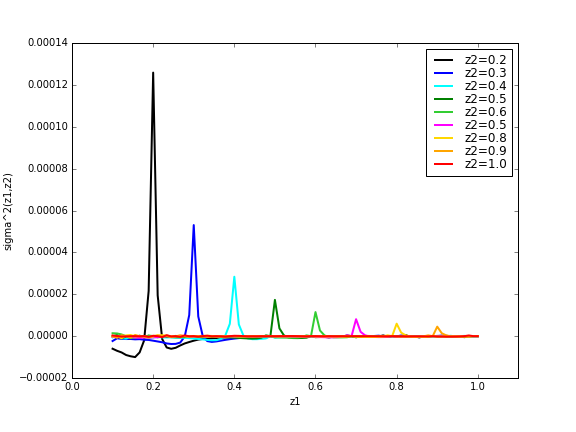}
\caption{Covariance of the background density: $\sigma^2_\mr{proj}(z_1,z_2)$ for different redshifts.}
\label{Fig:sigmaproj}
\end{figure}

Our framework also extends the work of \cite{Takada:2013wfa} (see their appendix A) 
since we find that the SSC covariance of projected observables involves an integral over \emph{two} redshifts instead of a single redshift. 
If we were able to use Limber's approximation, then $\sigma^2_\mr{proj}(z_1,z_2)$ would collapse to a Dirac delta function of redshifts and we would indeed end up with a single redshift integral. However, as is evident from figure \ref{Fig:sigmaproj}, this covariance does have a redshift width.

Physically, this means that the background density of two neighbouring slices of the past light cone are correlated. As the redshift separation increases, this correlation decreases with oscillations, asymptoting to zero. This behaviour is the reason why neighbouring redshift bins have some anti-correlation in figure \ref{Fig:Cov-Ncl}. 
We note that this behaviour would be even more important for finer redshift bins, e.g. for analysis of a spectroscopic or semi-spectroscopic survey.

Finally in appendix \ref{App:SSC-ineq}, we show that super-sample covariance obeys the following mathematical inequality:
\be
\label{Eq:SSC-ineq}\Cov_\mr{SSC}\left(\mathcal{O}_1 , \mathcal{O}_2\right) \leq \sqrt{\Cov_\mr{SSC}\left(\mathcal{O}_1 , \mathcal{O}_1\right) \times \Cov_\mr{SSC}\left(\mathcal{O}_2 , \mathcal{O}_2\right)}
\ee
for any observables $\mathcal{O}_1$ and $\mathcal{O}_2$.\\
As an example, for number counts this implies
\ba
\nonumber \Cov_\mr{2h}\left(N_\mr{cl}(i_M,i_z) , N_\mr{cl}(j_M,j_z)\right) &\leq \Big( \Cov_\mr{2h}\left(N_\mr{cl}(i_M,i_z) , N_\mr{cl}(i_M,i_z)\right) \\
& \times \Cov_\mr{2h}\left(N_\mr{cl}(j_M,j_z) , N_\mr{cl}(j_M,j_z)\right)\Big)^{1/2}
\ea
for any bins of redshifts and mass $i_z, j_z, i_M, j_M$.\\
In our case, we found inequality eq.~(\ref{Eq:SSC-ineq}) to be saturated by some terms of the joint covariance matrix which will be described in section \ref{Sect:jointcov}, and could even be sometimes violated at the percent level due to numerical errors. As such, enforcing inequality eq.~(\ref{Eq:SSC-ineq}) a posteriori would help to regularize the covariance matrix in cases where it was highly degenerate (low redshifts, small angular scales).

More generally we show in appendix \ref{App:SSC-ineq} that super-sample covariance is positive. That is, even in the absence of the other covariance terms, the covariance matrix is certain to have positive eigenvalues.

\section{Covariance of the galaxy angular power spectrum}\label{Sect:covgal}

Let us recall the equation for the galaxy power spectrum estimator eq.~(\ref{Eq:Clgal-estimator-Lagrange}):
\ba
\nonumber \hat{C}_\ell^\mr{gal}(i_z) &= 4\pi \int \dd V_{12} \; \frac{\overline{n}_\mr{gal}(z_1) \, \overline{n}_\mr{gal}(z_2)}{\Delta N_\mr{gal}(i_z)^2} \int \frac{\dd^3\kk_{12}}{(2\pi)^6} \; j_\ell(k_1 r_1) \,  j_\ell(k_2 r_2)\\
& \qquad \times P_\ell(\hk_1 \cdot \hk_2) \ \delta_\mr{gal}(\kk_1,z_1) \; \delta^*_\mr{gal}(\kk_2,z_2)
\ea
Taking the covariance of this estimator will involve the quantity 
\be\label{Eq:4ptavg-forcovcl}
\lbra \delta_\mr{gal}(\kk_1,z_1) \; \delta^*_\mr{gal}(\kk_2,z_2) \; \delta_\mr{gal}(\kk_3,z_3) \; \delta^*_\mr{gal}(\kk_4,z_4)\rbra
\ee
The unconnected part of this average yields the usual Gaussian $C_\ell$ covariance:
\ba
\Cov\left(\hat{C}_\ell^\mr{gal}(i_z),\hat{C}_{\ell'}^\mr{gal}(j_z)\right) &= \frac{2 \ C_\ell^\mr{gal}(i_z,j_z)^2}{2\ell+1} \ \delta_{\ell,\ell'}\\
&= \frac{2 \ C_\ell^\mr{gal}(i_z)^2}{2\ell+1} \ \delta_{\ell,\ell'} \ \delta_{i_z,j_z}
\ea
where the second equality is a consequence of Limber's approximation forcing $i_z=j_z$.\\
The connected part of eq.~(\ref{Eq:4ptavg-forcovcl}) is
\ba
\nonumber \lbra \delta_\mr{gal}(\kk_1,z_1) \, \delta^*_\mr{gal}(\kk_2,z_2) \, \delta_\mr{gal}(\kk_3,z_3) \, \delta^*_\mr{gal}(\kk_4,z_4)\rbra_c &= (2\pi)^3 \, \delta^{(3)}(\kk_1 - \kk_2 + \kk_3 -\kk_4) \\
& \qquad \times T_\mr{gal}(\kk_1,-\kk_2,\kk_3,-\kk_4)
\ea
where $T_\mr{gal}$ is the galaxy 3D trispectrum. Such term yields in principle the power spectrum covariance \cite{Lacasa2014b}
\be
\Cov\left(\hat{C}_\ell^\mr{gal}(i_z),\hat{C}_{\ell'}^\mr{gal}(j_z)\right) = \frac{1}{4\pi} \ T_{\ell \ell}^{\ell' \ell'}(\ell_\mr{diag}=0 | i_z,j_z)
\ee
where $T_{\ell_1 \ell_2}^{\ell_3 \ell_4}(\ell_\mr{diag})$ is the 2D (projected) trispectrum (see the derivation of eq.~(2.80) in chapter 2 of \cite{Lacasa2014b} for details).

The galaxy 3D trispectrum is a complicated beast with the contribution from many terms: 
there are 14 different diagrams, given e.g. in figure 7 of \cite{Lacasa2014a}, with non-trivial permutations and 
some diagrams splitting into contributions from perturbation theory, both at second and third order, from halo quadratic and cubic local bias as well as from non-local bias effects. 
A complete census of these terms is beyond the scope of this article and left to a future work. Furthermore, there is no analytical solution known to the authors for the 2D projection of a 3D trispectrum with a general momentum dependence. This projection is known to us only for the diagonal-independent case, i.e. $T_\mr{gal}$ depending only on the moduli $(k_1,k_2,k_3,k_4)$.
In that case, the projection is given in appendix E of \cite{Lacasa2014a} and recast here for convenience for the case of interest:
\be
T_{\ell \ell}^{\ell' \ell'}(\ell_\mr{diag} | i_z,j_z) = \delta_{i_z,j_z} \int \dd V \ \frac{\nbargal(z)^4}{\Delta N_\mr{gal}(i_z)^4} \ T_\mr{gal}(k_{\ell},k_{\ell},k_{\ell'},k_{\ell'} | z)
\ee
This is the case of the 1-halo term of the galaxy trispectrum, that is when the four points of the correlation function hit four different galaxies inside the same halo:
\ba
\nonumber T_\mr{gal}^{1h}(k_1,k_2,k_3,k_4 | z) &= \int \dd M \ \frac{\dd n_h}{\dd M} \ \frac{\lbra N(N-1)(N-2)(N-3)\rbra(M)}{\overline{n}_\mr{gal}^4}\\
& \qquad \times u(k_1|M,z) \; u(k_2|M,z) \; u(k_3|M,z) \; u(k_4|M,z).
\ea
We also found this to be the case of 1-halo shot-noise terms, that is when two or more points of the correlation function hit the same galaxies, and all galaxies are in the same halo. 
However we found these shot-noise terms to be negligible in all our tests.
\newline

An important covariance contribution which does not fit the previous criterion (diagonal-independence) is the super-sample covariance (SSC) discussed in section \ref{Sect:SSC}. By analogy with the SSC of the cross-covariance and following the discussion of section \ref{Sect:SSC}, we take:
\be
\Cov_\mr{SSC}\left(C_\ell^\mr{gal}(i_z),C_{\ell'}^\mr{gal}(j_z)\right) = \int \dd V_{12} \; \frac{\nbargal(z_1)^2 \, \nbargal(z_2)^2}{\Delta N_\mr{gal}(i_z)^2 \, \Delta N_\mr{gal}(j_z)^2} \; \frac{\partial P_\mr{gal}(k_\ell,z_1)}{\partial \delta_\mr{b}} \, \frac{\partial P_\mr{gal}(k_{\ell'},z_2)}{\partial \delta_\mr{b}} \; \sigma^2_\mr{proj}(z_1,z_2)
\ee
where all terms have been defined previously. We leave a rigorous derivation of this equation for a future work.

In realistic data analysis, multipoles of the power spectrum are binned together. The binning and its effect on the covariance is discussed in detail in appendix \ref{App:bin-multipoles}. 
This aspect is critical to produce realistic covariance matrices. Indeed, the Gaussian diagonal error bars decrease as $1/\sqrt{\Delta\ell}$ with binning, 
whereas other terms like, in particular SSC, decrease much slower because of their large off-diagonal contribution.

\begin{figure}[htb]
\centering
\includegraphics[width=.8\linewidth]{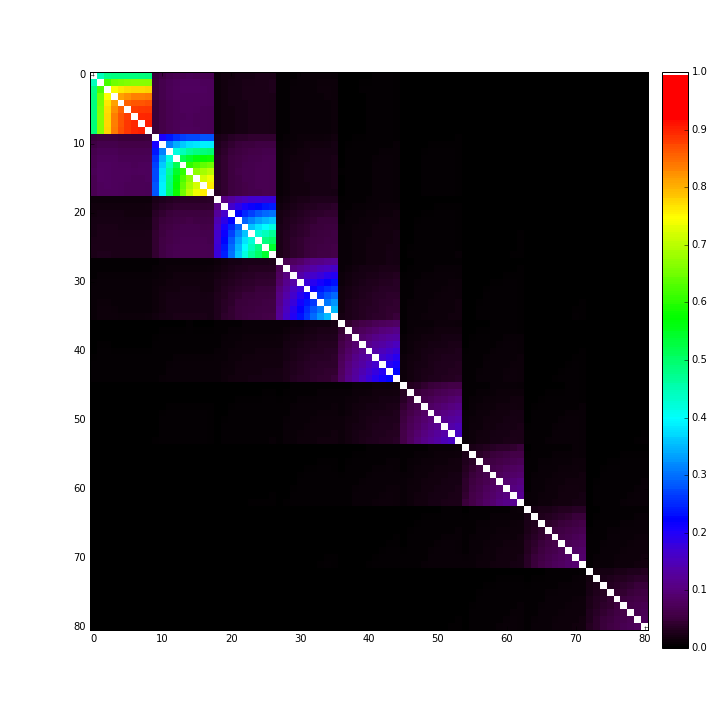}
\caption{Covariance of the galaxy angular power spectrum for nine multipole bins and nine redshift bins. Similar to figure \ref{Fig:Cov-Ncl} but for the galaxy spectrum instead of cluster counts. See text for details.}
\label{Fig:Cov-Clgal}
\end{figure}

Figure \ref{Fig:Cov-Clgal} shows the resulting correlation matrix (in absolute value) of the binned galaxy spectrum on large scales $\ell=30-300$ with 9  bins of width $\Delta\ell=30$ and
9 redshift bins of width $\Delta z = 0.1$ starting at $z=0.1$. Small scales will also be considered in the next section.
We see  that the covariance matrix of the galaxy spectrum is highly off-diagonal at low redshift, and still has $\sim 10\%$ off-diagonal elements in the highest redshift bin. Due to super-sample covariance, there is also some negative covariance between the neighbouring redshift bins at low redshift, which is plotted in absolute value in figure \ref{Fig:Cov-Clgal} and reaches $\sim 10\%$ for $0.1< z < 0.3$.

\section{Joint covariance and consequences for parameter constraints}\label{Sect:jointcov}

This section presents the main results of the article, showing the outcome of our modelling of the joint covariance of cluster counts and the galaxy angular power spectrum developed in the previous sections. 
The different contributions to the covariance matrix are analysed and the dominance of non-gaussian effects at low redshifts and small scales is emphasized.
A Fisher matrix study is used to evaluate the importance of combining these observables in the determination of cosmological and HOD parameters using different binnings in the angular scale to reflect large scales (mostly sensitive to cosmological parameters) or small scales (mostly sensitive to HOD parameters).

\subsection{Joint covariance} \label{Sect:subsect-jointcov}

As before, we consider nine redshift bins of width $\Delta z=0.1$ between $z_\mr{min}=0.1$ and $z_\mr{max}=1$, and three logarithmic mass bins for the cluster counts of width $\Delta\log M=0.5$ between $M_\mr{min}=10^{14} \Msun$ and $M_\mr{max}=10^{15.5} \Msun$.\\
The angular multipoles are binned in order to follow the more realistic procedure performed on real data.
The issue and the effects of binning are explained in more details in Appendix \ref{App:bin-multipoles}. We consider two case studies for the binning scheme:
\begin{itemize}
\item a ``cosmological'' case focusing on large angular scales, with nine regular multipole bins between $\ell_\mr{min}=30$ and $\ell_\mr{max}=300$ (9 bins of width $\Delta\ell=30$);
\item a ``HOD'' case focusing on smaller angular scales, with seven regular multipole bins between $\ell_\mr{min}=300$ and $\ell_\mr{max}=1000$ (7 bins of width $\Delta\ell=100$).
\end{itemize}
As implied by the names, the first case is expected to be representative of a cosmological constraint by focusing on linear scales, while the second is expected to be representative of a constraint on the Halo Occupation Distribution parameters by focusing on small non-linear scales.
It is not, however, the purpose of this article to study rigorously the separation of the linear and non-linear regimes. The inclusion of external priors and a detailed investigation of a cosmological study are left for future work.

\begin{figure}[htb]
\centering
\includegraphics[width=.49\linewidth]{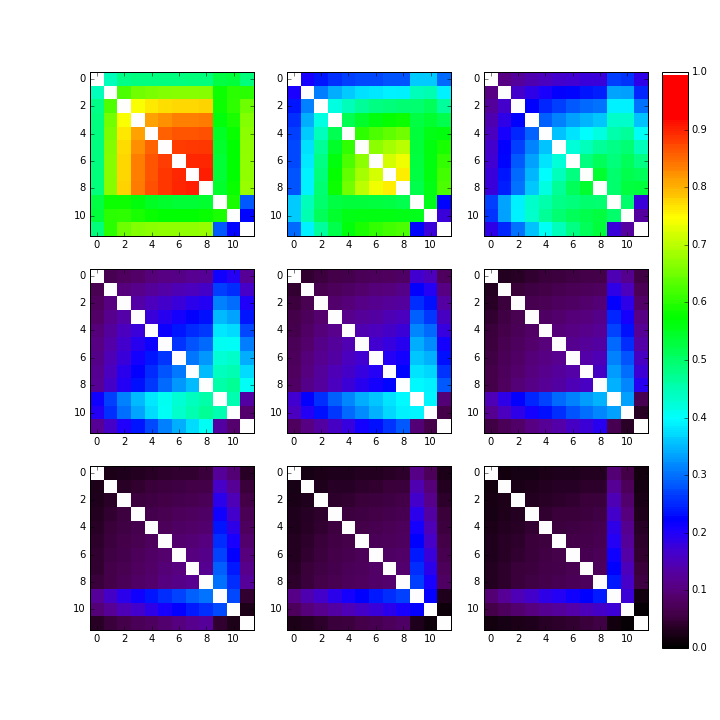}
\includegraphics[width=.49\linewidth]{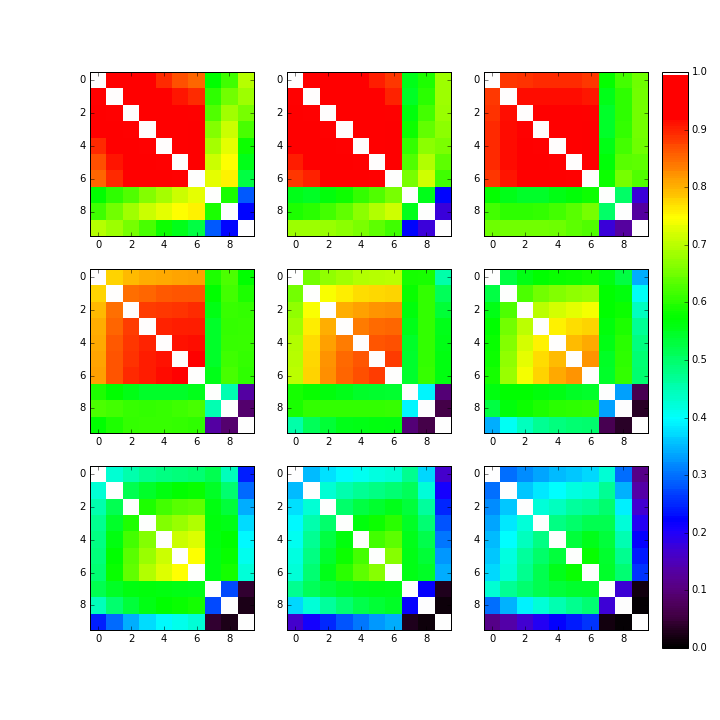}
\caption{Joint correlation matrix of $(C_\ell^\mr{gal},N_\mr{cl})$ in nine redshift shells, in the cosmological case (left panel) and in the HOD case (right panel). The auto-covariance of $C_\ell^\mr{gal}$ is the upper left block (respectively 9x9 in the cosmological case and 7x7 in the HOD case), the auto-covariance of $N_\mr{cl}$ is the 3x3 bottom right block, and the cross-covariance is given in the rectangular blocks. See text for details.}
\label{Fig:joint-cov}
\end{figure}

The joint correlation matrix of the galaxy power spectrum and the cluster counts can be seen in figure \ref{Fig:joint-cov}, with the cosmological case on the left panel and the HOD case on 
the right panel. 

This figure is one of the main results of this article.
One can clearly see the importance of non-Gaussian off-diagonal terms at low redshift. As expected, their significance decreases at higher redshifts, as perturbations become smaller 
and linear theory becomes a better approximation. Note however that even at redshift $z=0.9-1$ and large angular scales (left panel), the off-diagonal elements reach $\geq 20 \%$ in the cross-covariance, showing how important it is to account for non-Gaussianity.

Furthermore, one sees that non-Gaussianity becomes dramatic at small angular scales (right panel), to the point that it reaches total and irremediable domination at low redshift. 
Properly accounting for these effects is thus absolutely critical to analyse the small scale power spectrum, as they saturate the information content of the galaxy power spectrum alone.

\begin{figure}[htb]
\centering
\includegraphics[width=.32\linewidth]{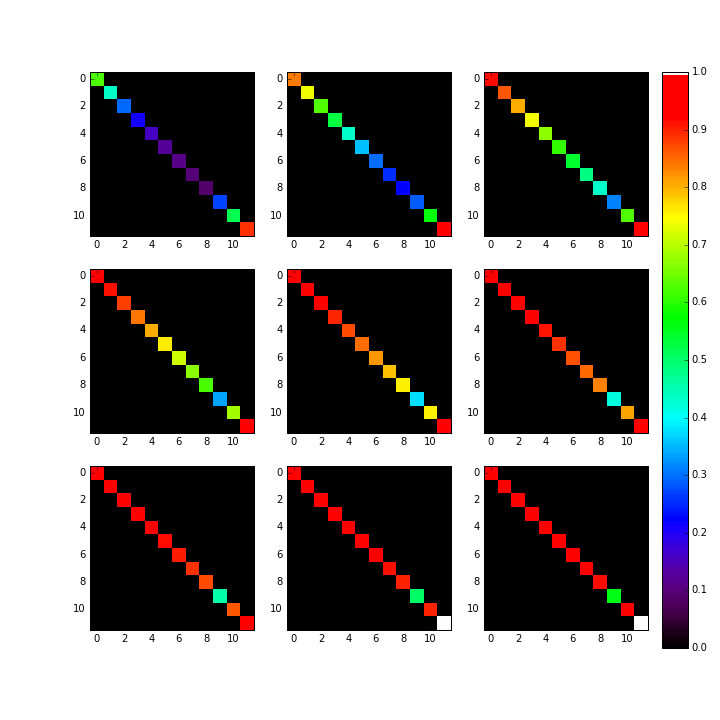}
\includegraphics[width=.32\linewidth]{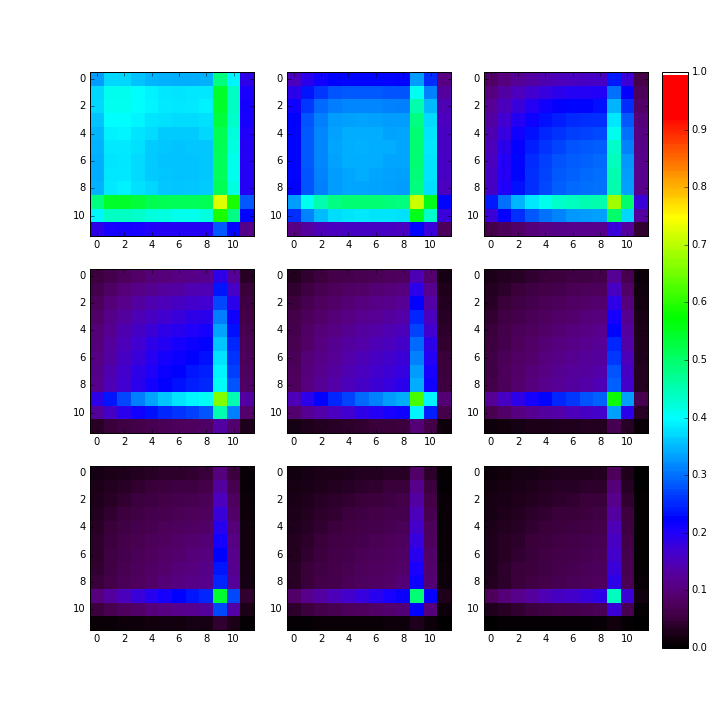}
\includegraphics[width=.32\linewidth]{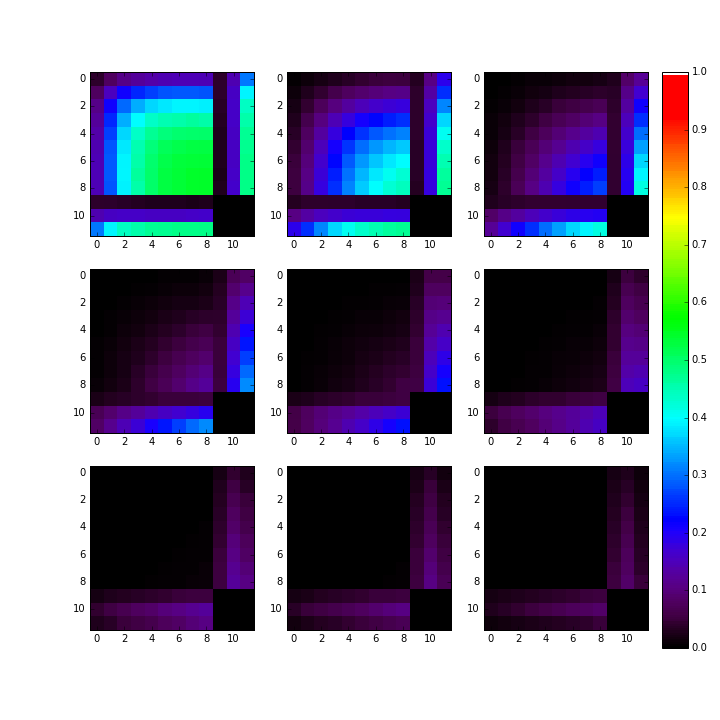}
\caption{Separate contributions to the joint covariance matrix in the cosmological case: standard terms (left panel), super-sample covariance (middle panel) and 
other non-Gaussian terms (right panel). All matrix elements are normalised to the \emph{total} variance, so that the sum of these three matrices gives the left panel of Fig. \ref{Fig:joint-cov}.}
\label{Fig:joint-cov-splitup}
\end{figure}

To make more evident the different contributions to the covariance matrix arising from standard terms (Gaussian for $C_\ell^\mr{gal}$ and Poissonian for $N_\mr{cl}$), 
super-sample covariance and other non-Gaussian terms, we show in figure \ref{Fig:joint-cov-splitup} their separate contributions in the cosmological (large scales) case.
We see that standard covariance terms only contribute to the diagonal of the covariance matrix, as expected, and that they dominate at high redshift. As redshift decreases and the universe gets more structured, their relative contribution decreases and they end up being important only for $C_\ell^\mr{gal}$ on the largest scales and for $N_\mr{cl}$ at the highest mass bin.

Super-sample covariance gives a relatively smooth contribution across the elements of the covariance matrix, 
being higher at low redshift though still reaching 10-20\% at high redshift. For the cross-covariance, it correlates the cluster counts at the lowest mass 
with all the galaxy spectrum, though more importantly with the smallest scales.

The other non-Gaussian terms have a more complex behaviour, with strong dependence on scale, redshift and mass. It is mostly negligible at high redshifts, especially in the auto-covariances, but become very important ar small redshifts. It increases strongly with multipole as the non-linear part of the galaxy spectrum becomes important. For the cross-covariance, it correlates mostly the cluster counts at the highest masses with the galaxy spectrum on small scales, even reaching large scales in the lowest redshifts.

One measure of the importance of the probes is the so-called signal-to-noise ratio (SNR). 
The SNR of an observable $\mathcal{O}$ with covariance matrix $\mathcal{C}$ is computed through the standard equation:
\be
\left(SNR\right)^2 = \mathcal{O}^T \cdot \mathcal{C}^ {-1} \cdot \mathcal{O}.
\ee
Using our computed covariance matrices in the cosmological and HOD cases, 
we show in  table \ref{Tab:SNR} the SNR for each probe and the joint combination. 
\begin{table}[htb]
\centering
\begin{tabular}{c|c|c|c|c}
case / probe(s) & $C_\ell^ \mr{gal}$ & $N_\mr{cl}$ & joint & independent \\ 
\hline 
cosmological & 141.09 & 162.18 & 194.35 & 214.96 \\ 
\hline 
HOD & 161.55 & 171.63 & 235.70 & 237.61 \\
\end{tabular}
\caption{Signal-to-noise ratios of the different probes and their combination, in the two case studies.}
\label{Tab:SNR}
\end{table}

The two probes have comparable SNR, and as expected the combined analysis yields a larger value, being even higher in the hypothetical independent case than in the case
where the cross-covariance is properly accounted for. In the next subsections we present a more detailed analysis of this comparison.

\subsection{Fisher matrix constraints on cosmological parameters}\label{Sect:Fisher-cosmo}

In this subsection, we compare the strength of cosmological parameter constraints that can be reached by using the galaxy angular power spectrum, the cluster number counts, and their combination. We also consider the hypothetical case where the galaxy spectrum and the number counts are independent, in order to isolate the particular effect of properly 
including their cross-covariance.
We emphasize that these constraints are not meant to be completely realistic, as we did not include a number of experimental effects in the modelling developed here 
(in particular photometric redshift errors, purity and completeness of cluster detections, and scatter in the cluster mass determination).  
That said, the choices of redshift binning, mass binning and mass range that we made are sufficiently conservative to mitigate some experimental errors, 
so that we expect the inclusion of experimental effects to be a correction to our results that will not change qualitatively our conclusions.

The strength of parameter constraints are evaluated by a standard Fisher matrix formalism. Specifically for a vectorized observable $\mathcal{O}$ and model parameters $\theta_i$, the element $(i,j)$ of the Fisher matrix is given by:
\be
F_{i,j} = \left(\frac{\partial \mathcal{O}}{\partial \theta_i}\right)^T \cdot \mathcal{C}^ {-1} \cdot \frac{\partial \mathcal{O}}{\partial \theta_j}
\ee
and $F^{-1}$ is the covariance matrix of the Gaussian likelihood of parameters.

In this article, the parameter of interest are both cosmological $(\sigma_8,\Omega_m h^2,w_\mr{DE})$ and HOD parameters $(\alpha_\mr{sat},M_\mr{min},M_\mr{sat})$, with 
fixed $\sigma_{\log M} = 0.2$.
For this subsection we are concentrating on cosmological parameters by marginalising over HOD parameters without priors. 
In the framework of Fisher matrices, through properties of Gaussian multivariate probabilities, this corresponds simply to extracting the appropriate block of the inverse Fisher matrix. Given
\be\label{Eq:invFish-cosmoNhod}
F^{-1} = \left(
\begin{array}{cc}
F^{-1}_\mr{c,c} & F^{-1}_\mr{c,H} \\
F^{-1}_\mr{H,c} & F^{-1}_\mr{H,H}
\end{array}
\right)
\ee
where c stands for cosmology and H for HOD, $F^{-1}_\mr{c,c}$ is the covariance matrix of cosmological parameters after HOD marginalisation.
\footnote{It turns out to be important to marginalise over HOD parameters, instead of fixing them to their fiducial values, as that can significantly increase the cosmological error bars. An important example is $\sigma_8$ for which the error bars are increased by a factor larger than 6 with $C_\ell^\mr{gal}$ in the cosmological study case, due to its high degeneracy with galaxy bias.}

\begin{figure}[htb]
\centering
\includegraphics[width=1.\linewidth]{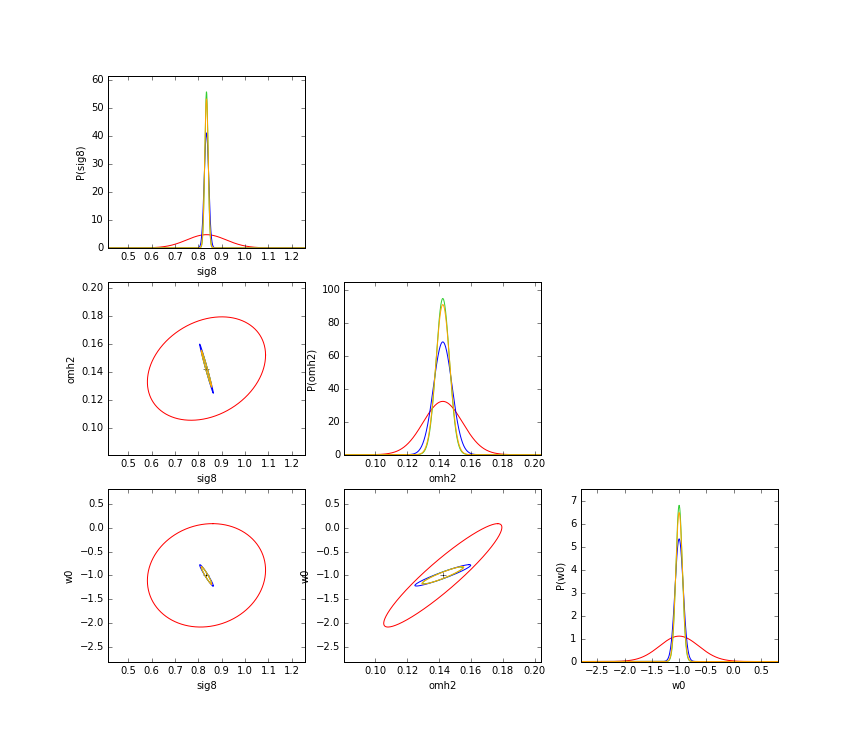}
\caption{1-dimensional pdfs and 3$\sigma$ 2-dimensional contours for cosmological parameter constraints in the cosmological study case. Red lines give the constraints from $C_\ell^\mr{gal}$, blue from $N_\mr{cl}$, orange from the joint analysis, green from the hypothetical independent case.}
\label{Fig:Fisher-cosmo}
\end{figure}

With this setup, figure \ref{Fig:Fisher-cosmo} shows the constraint on cosmological parameters in the cosmological case explained in section \ref{Sect:subsect-jointcov}
In this figure, the red lines give the Fisher ellipses and 1-dimensional probability distribution function (pdf) for the constraints coming from the galaxy power spectrum alone, 
the blue lines are for the constraints from the cluster number counts alone, orange is for the joint analysis, and finally green corresponds to the hypothetical case where 
the two probes were independent (i.e. setting the cross-covariance to zero).

Table \ref{Tab:errbar-cosmo} summarizes the marginalised error bars on each parameter that may be achieved in each case.
\begin{table}[htb]
\centering
\begin{tabular}{c|c|c|c}
observable/parameter & $\sigma_8$ & $\Omega_m h^2$ & $w_\mr{DE}$ \\ 
\hline 
$C_\ell^\mr{gal}$ & 10.1 \% & 8.67 \% & 36.1 \% \\ 
\hline 
$N_\mr{cl}$ & 1.17 \% & 4.10 \% & 7.48 \% \\ 
\hline 
joint & 0.90 \% & 3.08 \% & 6.16 \% \\ 
\hline
independent & 0.86 \% & 2.96 \% & 5.87 \% \\ 
\end{tabular}
\caption{1$\sigma$ marginalised error bars on cosmological parameters in the cosmological case study.}
\label{Tab:errbar-cosmo}
\end{table}

From figure \ref{Fig:Fisher-cosmo} and table \ref{Tab:errbar-cosmo}, one sees that constraints from the galaxy angular power spectrum are weaker 
than cluster counts constraints for the case studied. 
We emphasize that this is partly due to the choice of angular scales considered and that a study being able to reliably push the cosmological analysis to 
smaller scales would find better galaxy power spectrum constraints. 
In our case though, the difference of constraints between the galaxy spectrum and cluster counts is partly due to the small difference in SNR (see table \ref{Tab:SNR}), 
the degeneracy with HOD parameters, and the higher sensitivity of cluster counts to cosmology. 
In the case of $\sigma_8$, for instance, the galaxy constraints are weakened by a factor larger than 6 due to the degeneracy with galaxy bias, while it is well-known that cluster abundances are exceptionally sensitive to $\sigma_8$.

Due to the higher sensitivity of cluster counts to cosmology as compared to the galaxy angular power spectrum, 
the joint cosmological constraints are close to the cluster counts constraints, as can be seen in figure \ref{Fig:Fisher-cosmo}. An improvement of 17-25\% is still to be noted. 
Additionally, we see that the joint constraints are slightly worse than in the hypothetical independent case, so the inclusion of the cross-covariance slightly degrades the constraints.


\subsection{Fisher constraints on HOD parameters}\label{Sect:Fisher-HOD}

This subsection mirrors section~\ref{Sect:Fisher-cosmo}, but focusing on HOD parameters instead of cosmological ones. We follow the same Fisher equations, and HOD constraints are achieved after marginalisation over cosmology, which corresponds to extracting the block $F^{-1}_\mr{H,H}$ in eq.~(\ref{Eq:invFish-cosmoNhod}).\\
Contrary to section \ref{Sect:Fisher-cosmo} however, number counts alone do not produce any parameter constraints, as by definition cluster counts do not depend on HOD. 
Number counts however help in a combination with the galaxy angular power spectrum in two ways: 
first they reduce the error on cosmology, which helps when marginalising over it, and second they break the degeneracy between the mass function and HOD parameters, by fixing the former.

\begin{figure}[htb]
\centering
\includegraphics[width=1.\linewidth]{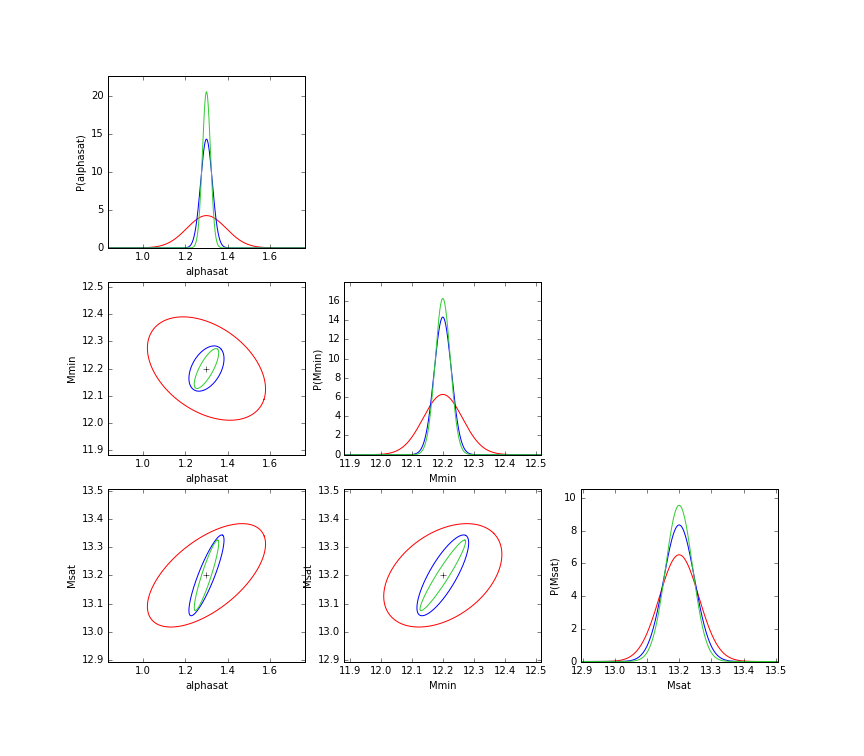}
\caption{1-dimensional pdfs and 3$\sigma$ 2-dimensional contours for HOD parameter constraints in the HOD study case. Red lines give the constraints from $C_\ell^\mr{gal}$, green from the joint analysis, blue from the hypothetical independent case.}
\label{Fig:Fisher-HOD}
\end{figure}

Figure \ref{Fig:Fisher-HOD} shows the forecasted constraints on HOD parameters, in the HOD case study. 
Red lines show the constraints from the galaxy power spectrum alone, green lines for the joint analysis and blue corresponds to the hypothetical independent case.

A quantitative assessment of the constraints is given in table \ref{Tab:errbar-HOD} which lists the marginalised $1\sigma$ error bars on each HOD parameter in the different cases.
%
\begin{table}[htb]
\centering
\begin{tabular}{c|c|c|c}
observable / parameter & $\alpha_\mr{sat}$ & $\log M_\mr{min}$ & $\log M_\mr{sat}$ \\ 
\hline 
$C_\ell^\mr{gal}$ & 7.20 \% & 0.52 \% & 0.46 \% \\ 
\hline 
joint & 1.49 \% & 0.20 \% & 0.32 \% \\ 
\hline
independent & 2.14 \% & 0.23 \% & 0.36 \% \\ 
\end{tabular}
\caption{1$\sigma$ marginalised error bars on HOD parameters in the HOD case study.}
\label{Tab:errbar-HOD}
\end{table}

From figure~\ref{Fig:Fisher-HOD} and table~\ref{Tab:errbar-HOD}, one sees that the galaxy HOD constraints are clearly improved by including the information from cluster counts. 
The improvement is the most spectacular for $\alpha_\mr{sat}$ where the errors bars are decreased by a factor larger than 4. This can be understood through the following reasoning. 
Since $\alpha_\mr{sat}$ governs the number of galaxies in massive haloes and massive haloes are rare, there can be important variations in their number due to cosmic variance. The effect of $\alpha_\mr{sat}$ on the galaxy power spectrum is thus highly degenerate with the mass function. Hence including cluster counts measurement breaks this degeneracy and allows for a much stronger $\alpha_\mr{sat}$ constraint.

Furthermore, the parameter constraints are this time improved in the combined realistic case compared to the naive independent hypothesis, despite the joint probe having a lower 
SNR (see table \ref{Tab:SNR}). 
This is the opposite of the results on the cosmological constraints in the cosmology case, as reported in section~\ref{Sect:Fisher-cosmo} \footnote{In fact we checked that the improvement scenario (i.e. constraints being better in the realistic case than in the hypothetical independent case) is the most frequent. For instance, HOD constraints are also improved in the cosmological study case and the same goes for cosmological constraints in the HOD study case. However we refrain from reporting those results as they are unrealistic for actual data analysis.}.
This is evidence for a synergy of the probes and the breaking of degeneracies. Indeed, in the independent hypothesis, the increased HOD constraints, compared to $C_\ell^\mr{gal}$ alone, 
come only from the improvement of cosmological constraints which are then marginalised over.



\section{Conclusion}\label{Sect:conclusion}

With the advent of large surveys such as the Dark Energy Survey, which provides us with a rich number of different observables, comes the issue of combining probes to obtain stronger constraints. In this article, we developed the framework to combine two particular probes of galaxy surveys: the cluster number counts and the galaxy angular power spectrum. We worked in full sky and did not consider a number of experimental effects such as photometric redshift errors of galaxies and clusters, purity and completeness of cluster detections, and scatter in the cluster mass determination. However, we chose in our analysis conservative values for the width of redshift and mass bins that should mitigate the impact of these experimental limitations. We thus expect the inclusion of these effects to be a correction to our results, and hence it should not change qualitatively our conclusions.

We used the halo model and Halo Occupation Distribution to model both the cluster counts, the galaxy spectrum, their respective auto-covariances, and for the first time the cross-covariance between these two observables, accounting for the important non-Gaussian contributions to the covariances. In particular we took into account second order halo bias, galaxy shot noise, and showed that halo non-local bias gives a null contribution to the cross-covariance. We also discussed in details the case of the super-sample covariance and showed that it obeys general mathematical inequalities whatever the probes considered.

We presented the joint covariance matrix of these observables for two cases focussing either on large or small scales. In both cases we pointed out the importance of non-Gaussian contribution to the covariance matrix, which become irremediably dominant at low redshifts and/or small scales. A careful modeling of non-linearities is thus essential if one wants to extract useful information in this regime. Furthermore we showed that a fully non-linear model is required to reproduce the cross-covariance in the full range needed, as perturbation theory alone fails on small scales, low redshifts and/or high cluster masses. A non-perturbative model is thus essential to combine these two probes.

We then used our predicted covariance matrix to analyse the consequences of combining both observables, showing first the increase of signal-to-noise with respect to the single probe case. Next we conducted a Fisher matrix analysis in the two cases : first concentrating on large scales to constrain cosmological parameters while marginalising over HOD parameters, and second focussing on small scales to constrain HOD parameters while marginalising over cosmological parameters.
For cosmology, compared to the best single-probe constraints, improvements of the order of $20 \%$ can be expected by including the combination. For HOD, although it is not constrained by cluster counts alone, the inclusion of cluster counts breaks the degeneracy between the mass function and HOD. This results in a dramatic improvement of the bounds on HOD parameters, reducing the error bars by a factor of 1.4 to 4.8 depending on the parameter.\\
These results demonstrate the critical importance of combining these observables in deriving constraints on cosmology and HOD.

Finally, compared to an hypothetical independent case, we showed that the consistent inclusion of the cross-correlation of the probes can lead to a few percent degradation in the cosmological constraints, but it leads to a large improvement of the HOD constraints, up to $50 \%$ in the case of $\alpha_\mr{sat}$. This shows conclusively the importance of accounting for the cross-covariance in the combination of these probes.


\appendix
\section*{Appendix}
\section{3D halo-galaxy-galaxy bispectrum in the Halo Model}\label{App:3Dhgg}
The different terms of the halo-galaxy-galaxy bispectrum can be derived from eq.~(\ref{Eq:halodensity-sumi}) \& eq.~(\ref{Eq:galdensity-sumij}) through lengthy and error-prone calculations. Alternatively, \cite{Lacasa2014a} have shown that the galaxy polyspectra equations can be derived via an elegant diagrammatic formalism with `Feynman'-like rules. Here we follow this approach and extend it to cross-polyspectra with halo statistics in given mass bins.\\
For the halo-galaxy-galaxy bispectrum, the first step is to draw all the diagrams involving one halo for the cluster count, labeled "1", and two possibly merged galaxies for the galaxy spectrum, labeled "2" and "3". Merging occurs for shot-noise diagrams. These diagrams are shown in Fig. \ref{Fig:diagrams}.\\
Then a diagram contributes as a term to the hgg bispectrum given by:
\begin{itemize}
\setlength{\itemsep}{0pt}
\item a prefactor $1 / (\nbargal(z_2) \, \nbargal(z_3))$
\item a possible Kronecker of redshifts if some halos coincide, e.g. $\delta_{z_1,z_2}$ if the first galaxy (labeled "2") is in the halo of the number count (labeled "1").
\item integrals over the halo mass $\int \dd M \ \frac{\dd n_h}{\dd M}$ for the halos containing the galaxies, except if that is the halo of the number count (since that one has a fixed mass $M_1$).
\item average number of uplet of galaxy given by the HOD, e.g. $\lbra N_\mr{gal}(M)\rbra$ for a single galaxy in a halo, $\lbra N(N-1)\rbra$ for a pair.
\item for each galaxy, halo profile $u(k|M)$ at the mass of the host halo, with k being the norm of the sum of Fourier wavevectors over the galaxy labels, i.e. if the galaxy has label "2" $k=k_2$, but for a shot-noise diagram $k = |\kk_2 + \kk_3| = k_1$
\item the halo polyspectrum of the considered halos, conditionned to their mass and redshift
\item a possible permutation of indices to account for symmetries. e.g. in the 2h-2h diagram of Fig. \ref{Fig:diagrams}, we have put galaxy "2" in the halo "1" of counts, but there is the symmetric diagram where galaxy "3" is in the halo "1".
\end{itemize}
Note that this diagrammatic formalism and these rules allow to compute any polyspectrum of halos and galaxies, though we just need the hgg bispectrum in this article.
We note that the complexity increases very quickly with the polyspectrum order considered, with e.g. the hggg trispectrum requiring 14 different terms, so that this diagrammatic becomes increasingly useful with order to avoid lengthy and cumbersome calculations.

Applying this diagrammatic formalism to the hgg bispectrum, we get 6 different diagrams/terms with the following expressions:
\ba
\nonumber B^{3h}(k_{123},z_{123} | M_1,z_1) &= \frac{1}{\overline{n}_\mr{gal}(z_2) \, \overline{n}_\mr{gal}(z_3)} \int \dd M_{23} \left.\frac{\dd^2 n_h}{\dd M}\right|_{M_2,z_2} \left.\frac{\dd^2 n_h}{\dd M}\right|_{M_3,z_3} \\
\nonumber & \qquad \qquad \times  \lbra N_\mr{gal}(M_2) \rbra \, \lbra N_\mr{gal}(M_3) \rbra \, u(k_2 | M_2) \, u(k_3 | M_3) \\
\label{Eq:B3h} & \qquad \qquad \qquad \qquad \times B_\mr{halo}(k_{123} | M_{123}, z_{123})\\
\nonumber B^{2h-2h}(k_{123},z_{123} | M_1,z_1) &= \frac{\delta_{z_1,z_2}}{\overline{n}_\mr{gal}(z_2) \, \overline{n}_\mr{gal}(z_3)} \int \dd M_3 \left.\frac{\dd^2 n_h}{\dd M}\right|_{M_3,z_3} \!\!\! \lbra N_\mr{gal}(M_1) \rbra \lbra N_\mr{gal}(M_3) \rbra\\
\label{Eq:B2h2h} &  \qquad \times u(k_2|M_1) \, u(k_3|M_3) \; P_\mr{halo}(k_3 \, | M_{13},z_{13}) \ + \mr{perm.\, (2\leftrightarrow 3)}\\
\nonumber B^{2h-1h2g}(k_{123},z_{123} | M_1,z_1) &= \frac{\delta_{z_2,z_3}}{\overline{n}_\mr{gal}(z_2) \, \overline{n}_\mr{gal}(z_3)} \int \dd M \, \frac{\dd^2 n_\mr{h}}{\dd M} \, \lbra N_\mr{gal}(N_\mr{gal}-1)(M) \rbra \\
\label{Eq:B2h-1h2g} & \qquad \times u(k_2 | M) \, u(k_3 | M) \; P_\mr{halo}(k_1 | M_1,M,z_1,z_2)\\
\nonumber B^{2h-1h1g}(k_{123},z_{123} | M_1,z_1) &= \frac{\delta_{z_2,z_3}}{\overline{n}_\mr{gal}(z_2) \, \overline{n}_\mr{gal}(z_3)} \int \dd M \, \frac{\dd^2 n_\mr{h}}{\dd M} \, \lbra N_\mr{gal}(M) \rbra \\
\label{Eq:B2h-1h1g} & \qquad \times u(k_1 \, | M) \, P_\mr{halo}(k_1 | M_1,M,z_1,z_2)\\
\label{Eq:B1h2g} B^{1h2g}(k_{123},z_{123} | M_1,z_1) &= \frac{\delta_{z_1,z_2,z_3}}{\overline{n}_\mr{gal}(z_2) \, \overline{n}_\mr{gal}(z_3)} \lbra N_\mr{gal}(N_\mr{gal}-1)(M_1) \rbra u(k_2 | M_1) \, u(k_3 | M_1)\\
\label{Eq:B1h1g} B^{1h1g}(k_{123},z_{123} | M_1,z_1) &= \frac{\delta_{z_1,z_2,z_3}}{\overline{n}_\mr{gal}(z_2) \, \overline{n}_\mr{gal}(z_3)} \lbra N_\mr{gal}(M_1) \rbra \, u(k_1 | M_1)
\ea
where $\delta_{z_1,z_2,z_3}=\delta_{z_1,z_2} \; \delta_{z_2,z_3}$.\\
Note that the $2h-1h1g$ and $1h1g$ terms are shot-noise terms and are thus irrelevant for the covariance with the real-space ACF (as shot-noise, corresponding to $\theta=0$, is explicitly not considered), but they are relevant for the covariance with the power spectrum.
\newline

Eq. \ref{Eq:B3h} involves the halo bispectrum, which we compute at tree-level using the following equation for the halo density:
\be
\delta_h(\xx|M,z) = b_1(M,z) \, \delta_\mr{DM}(\xx,z) + \frac{b_2(M,z)}{2!} \, \left(\delta_\mr{DM}(\xx,z)^2 - \lbra \delta_\mr{DM}^2\rbra\right) + b_{s2}(M,z) \, \left(s^2(\xx,z) - \lbra s^2 \rbra \right)
\ee
where $s^2$ is the square of the tidal tensor and we have \cite{Baldauf2012}:
\be
s^ 2(\kk,z) = \int \frac{\dd \kk'}{(2\pi)^ 3} \, S_2(\kk',\kk-\kk') \; \delta_\mr{DM}(\kk',z) \, \delta_\mr{DM}(\kk-\kk',z) \quad \mr{with} \quad S_2(\kk_1,\kk_2) = \frac{(\kk_1 \cdot \kk_2)^2}{k_1^2 \, k_2^2} - \frac{1}{3}
\ee
The resulting halo bispectrum has the expression~:
\ba
\nonumber B_\mr{halo}(k_{123} | M_{123}, z_{123}) &= b_1(M_1,z_1) \, b_1(M_2,z_2) \, b_1(M_3,z_3) \, B_\mr{DM}(k_{123} , z_{123}) \\
\nonumber & \quad + \big[ b_2(M_1,z_1) \, b_1(M_2,z_2) \, b_1(M_3,z_3) \\
\nonumber & \qquad \times P_\mr{DM}(k_2 | z_{12}) \, P_\mr{DM}(k_3 | z_{13}) + 2 \ \mr{perm.} \big]\\
\nonumber & \quad + \big[ 2 \, b_{s2}(M_1,z_1) \, b_1(M_2,z_2) \, b_1(M_3,z_3) \\
& \qquad \times S_2(\kk_2,\kk_3) \; P_\mr{DM}(k_2 | z_{12}) \, P_\mr{DM}(k_3 | z_{13}) + 2 \ \mr{perm.} \big]
\ea
where the dark matter bispectrum is given at second-order in perturbation theory :
\ba
B_\mr{DM}(k_{123} , z_{123}) &= 2 F^\mr{2PT}(\kk_1,\kk_2) \, P_\mr{lin}(k_1 | z_1, z_3) \, P_\mr{lin}(k_2 | z_2, z_3) + 2 \ \mr{perm.}
\ea
(adapted from \cite{Fry1984} for different redshifts)\\
with
\be
F^\mr{2PT}(\kk_\alpha,\kk_\beta) = 
\frac{5}{7} +\frac{1}{2} \cos\theta_{\alpha \beta} \, \left(\frac{k_\alpha}{k_\beta} + \frac{k_\beta}{k_\alpha}\right) + \frac{2}{7} \cos^2\theta_{\alpha \beta}.
\ee

\section{From the 3D hgg bispectrum to the cross-covariance}\label{App:crosscov-deriv}
From Eq. \ref{Eq:cov-Ncl-Clgal-estim}, the definition of the halo density contrast and the halo-galaxy-galaxy (hgg) bispectrum comes:
\ba \label{Eq:cov-ncl-clgal-orig}
\nonumber \Cov\left(N_\mr{cl}(i_M,i_z) , C_\ell^\mr{gal}(j_z)\right) &= 4\pi \int \dd M_1 \, \dd V_{123} \; \frac{\nbargal(z_2) \, \nbargal(z_3)}{\Delta N_\mr{gal}(j_z)^2} \; \left.\frac{\dd n_h}{\dd M}\right|_{M_1,z_1} \\
\nonumber & \times \int \frac{\dd^3\kk_{123}}{(2\pi)^9} \, j_0(k_1 r_1) \, j_\ell(k_2 r_2) \, j_\ell(k_3 r_3) \; P_\ell(\hk_2\cdot\hk_3) \\
& \quad \times (2\pi)^3 \, \delta^3(\kk_1+\kk_2-\kk_3) \, B_\mr{hgg}(\kk_{123}|M_1,z_{123})
\ea
We note that this can also be written as:
\ba \label{Eq:cov-ncl-clgal-rewritten}
\Cov\left(N_\mr{cl}(i_M,i_z) , C_\ell^\mr{gal}(j_z)\right) &= 4\pi \int \dd M_1 \, \dd V_{123} \, \left.\frac{\dd n_h}{\dd M}\right|_{M_1,z_1} 
\; \frac{\nbargal(z_2) \, \nbargal(z_3)}{\Delta N_\mr{gal}(j_z)^2} \,  b^\mr{hgg}_{0\ell\ell}(M_1,z_{123})
\ea
where $b^\mr{hgg}$ is the halo-galaxy-galaxy angular bispectrum per comoving volumes and unit mass. This bispectrum is involved in the squeezed configuration $\ell_1 \ll \ell_2,\ell_3$, in fact the most squeezed possible : $\ell_1=0$. Hence we will have contributions in the integral only from triplets of Fourier modes satisfying $k_1 \ll k_2 \approx k_3$.

Now the thirteen integrals of Eq. \ref{Eq:cov-ncl-clgal-orig} are not numerically tractable and thus need to be simplified analytically. To shorten further equations, we are going to first concentrate on the simplification of $b^\mr{hgg}_{0\ell\ell}(M_1,z_{123})$ :
\ba
\nonumber b^\mr{hgg}_{0\ell\ell}(M_1,z_{123}) &=  \int \frac{\dd^3\kk_{123}}{(2\pi)^9} \, j_0(k_1 r_1) \, j_\ell(k_2 r_2) \, j_\ell(k_3 r_3) \; P_\ell(\hk_2\cdot\hk_3) \\
& \quad \times (2\pi)^3 \, \delta^3(\kk_1+\kk_2-\kk_3) \, B_\mr{hgg}(k_{123}|M_1,z_{123})
\ea
We are first going to complexify this equation by writing the Legendre polynomial as as sum of spherical harmonics with the addition theorem, introduce the Fourier representation of the dirac, and expand the resulting Fourier modes in terms of Bessel functions and spherical harmonics :
\ba
\nonumber b^\mr{hgg}_{0\ell\ell}(M_1,z_{123}) &= \int \frac{\dd^3\kk_{123}}{(2\pi)^9} \, j_0(k_1 r_1) \, j_\ell(k_2 r_2) \, j_\ell(k_3 r_3) \frac{4\pi}{2\ell+1} \sum_m Y^*_{\ell m}(\hk_2) \, Y_{\ell m}(\hk_3) \\
\nonumber & \times B_\mr{hgg}(\kk_{123}|M_1,z_{123}) \int \dd^3 \xx \, (4\pi)^3 \sum_{1,2,3} i^{\ell_1+\ell_2-\ell_3} \, j_{\ell_1}(k_1 x) j_{\ell_2}(k_2 x) j_{\ell_3}(k_3 x) \\
\label{Eq:b0ll-hgg-fullexpanded} & \qquad \times Y_1(\hk_1) \, Y_2(\hk_2) \, Y^*_3(\hk_3) \ Y^*_1(\hat{x}) \, Y^*_2(\hat{x}) \, Y_3(\hat{x}) 
\ea
which sure looks better.

Now we are faced with an alternative depending on the bispectrum term considered : either the bispectrum term only depends on the moduli $(k_1,k_2,k_3)$ (case of most bispectrum terms), or it also depends on the angles between the Fourier vectors $(\hk_1\cdot\hk_2,\hk_1\cdot\hk_3,\hk_2\cdot\hk_3)$ (case of the 2PT and s2 terms). We are going to treat these cases in separate subsections.

\subsection{Angle-independent bispectrum}\label{Sect:angle-indep-bisp}

We first note that the derivation below is essentially a copy of the derivation of Appendix E of \cite{Lacasa2014a} in the particular case of the bispectrum in the squeezed limit $0\ell\ell$, except that we will not apply a Limber approximation on $k_1$.

In the angle-independent case, $B_\mr{hgg}(\kk_{123}|M_1,z_{123}) = B_\mr{hgg}(k_{123}|M_1,z_{123})$. Inputting this in Eq. \ref{Eq:b0ll-hgg-fullexpanded}, we see that the integrals over the unit vectors $\hk_1,\hk_2,\hk_3,\hat{x}$ can be trivially performed as they only involve spherical harmonics. Specifically the integrals over $\hk_1,\hk_2,\hk_3$ force $(\ell_1,m_1)=(0,0)$, $(\ell_2,m_2)=(\ell,m)$ and $(\ell_3,m_3)=(\ell,m)$. Taking care that $Y_{00} = 1/\sqrt{4\pi}$, the integral over $\hat{x}, \hk_1$ and the sum over m are trivial and we get:
\ba
\nonumber b^\mr{hgg}_{0\ell\ell}(M_1,z_{123}) &= \frac{\nbargal(z_2) \, \nbargal(z_3)}{\Delta N_\mr{gal}(j_z)^2} \left(\frac{1}{2\pi^2}\right)^3 \int k^2_{123} \, \dd k_{123} \; 4\pi \, x^2 \, \dd x \; j_0(k_1 r_1) \, j_\ell(k_2 r_2) \, j_\ell(k_3 r_3) \\
& \qquad \times j_0(k_1 x) \, j_\ell(k_2 x) \, j_\ell(k_3 x) \; B_\mr{hgg}(k_{123}|M_1,z_{123})
\ea
We now apply the Limber approximation on $k_2$ and $k_3$, that is to assume that the bispectrum varies slowly with these arguments compared to the oscillations of the corresponding Bessel functions. We can thus take out the bispectrum of these integrals, evaluating it at the Bessel peak $(\ell+1/2)/r_{2/3}$, and use the identity:
\be
\int k^2 \, \dd k \; j_\ell(k r) \, j_\ell(k x) = \frac{\pi}{2 r^2} \delta(r-x)
\ee
This greatly simplifies the bispectrum:
\ba
\nonumber b^\mr{hgg}_{0\ell\ell}(M_1,z_{123}) &= \frac{\nbargal(z_2) \, \nbargal(z_3)}{\Delta N_\mr{gal}(j_z)^2} \frac{1}{(2\pi)^3} \int k^2_{1} \, \dd k_{1} \; x^2 \, \dd x \; j_0(k_1 r_1) \, j_0(k_1 x) \, B_\mr{hgg}(k_1,k_\ell,k_\ell|M_1,z_{123}) \\
& \qquad \times \frac{\delta(r_2-x) \ \delta(r_3-x)}{r_2^2 \ r_3^2}
\ea
where $k_\ell = (\ell+1/2)/r_{2}$ (since $r_2=r_3=x$).\\
Inputting this in Eq. \ref{Eq:cov-ncl-clgal-rewritten} for the covariance gives:
\ba
\nonumber \Cov\left(N_\mr{cl}(i_M,i_z) , C_\ell^\mr{gal}(j_z)\right) &= 4\pi \int \dd M_1 \, \dd V_{123} \, \dd V_x \, \left.\frac{\dd n_h}{\dd M}\right|_{M_1,z_1} \! \frac{\nbargal(z_2) \, \nbargal(z_3)}{\Delta N_\mr{gal}(j_z)^2} \, \frac{\delta(r_2-x) \ \delta(r_3-x)}{r_2^2 \ r_3^2} \\
& \times \frac{1}{(2\pi)^3} \int k^2 \, \dd k \; j_0(k r_1) \, j_0(k x) \, B_\mr{hgg}(k,k_\ell,k_\ell|M_1,z_{123}) \\
\nonumber &= \int \dd M_1 \, \dd V_{1} \, \dd V_x \, \left.\frac{\dd n_h}{\dd M}\right|_{M_1,z_1} \!\!\! \frac{\nbargal(z_x)^2}{\Delta N_\mr{gal}(j_z)^2} \\
\label{Eq:crosscov-proj-2redshifts} & \times \frac{1}{2 \pi^2} \int k^2 \, \dd k \; j_0(k r_1) \, j_0(k x) \, B_\mr{hgg}(k,k_\ell,k_\ell|M_1,z_1,z_x,z_x)
\ea
where $x$ has been considered has the comoving distance to a redshift $z_x$, $k_\ell = (\ell+1/2)/x$ and the integral over $z_x$ implicitly runs over the redshift bin $j_z$.\\
This equation allows to compute the cross-covariance from the 2h-1h1g, and 2h-1h2g bispectrum terms (Eq. \ref{Eq:B2h-1h1g}-\ref{Eq:B2h-1h2g}), giving explicitely:
\ba
\nonumber \Cov_\mr{2h-1h1g}\left(N_\mr{cl}(i_M,i_z) , C_\ell^\mr{gal}(j_z)\right) &= \frac{1}{\Delta N_\mr{gal}(j_z)^2} \int \dd V_{12} \, \frac{\partial n_h}{\partial \delta_b}(i_M,z_1) \\
& \times \nbargal(z_2) \, b_1^\mr{gal,eff}(z_2) \; \sigma^2_\mr{proj}(z_1,z_2) \\
\nonumber \Cov_\mr{2h-1h2g}\left(N_\mr{cl}(i_M,i_z) , C_\ell^\mr{gal}(j_z)\right) &= \frac{1}{\Delta N_\mr{gal}(j_z)^2} \int \dd V_{12} \, \frac{\partial n_h}{\partial \delta_b}(i_M,z_1) \\
& \times \nbargal(z_2) \, b_1^\mr{gal,eff-2}(k_\ell,z_2) \; \sigma^2_\mr{proj}(z_1,z_2)
\ea
where in the 2h-1h1g term, we approximated $u(k_1|M) \approx 1$, as $k_1$ gets aliased into the monopole. That is, no considered halo takes up the whole sky. This seems a fair approximation as none of our observables is sensitive to the local group.

Eq. \ref{Eq:crosscov-proj-2redshifts} also allow to compute the cross-covariance from the b2 subterm of the 3h bispectrum. Recalling
\be
B_\mr{halo}^\mr{b2}(k_{123}|M_{123},z_{123}) = b_2(M_1,z_1) \, b_1(M_2,z_2) \, b_1(M_3,z_3) \; P_\mr{DM}(k_2|z_{12}) \, P_\mr{DM}(k_3|z_{13}) + 2 \, \mr{perm.}
\ee
and given that $k_1 \ll k_2,k_3$ and that $P(k)$ is decreasing over the range of interest, we can assume that $P(k_2) P(k_3) \ll P(k_1) P(k_2) , P(k_1) P(k_3)$ and neglect the corresponding term. This yields:
\ba
\nonumber \Cov_\mr{b2}\left(N_\mr{cl}(i_M,i_z) , C_\ell^\mr{gal}(j_z)\right) &= \frac{2}{\Delta N_\mr{gal}(j_z)^2} \int \dd V_{12} \; \frac{\partial n_h}{\partial \delta_b}(i_M,z_1) \; \nbargal(z_2)^2\\
& \times b_2^\mr{gal,eff}(k_\ell,z_2) \, b_1^\mr{gal,eff}(k_\ell,z_2) P_\mr{DM}(k_\ell | z_2) \; \sigma^2_\mr{proj}(z_1,z_2) 
\ea

Some bispectrum terms do not depend on $k_1$ (2h-2h and 1h2g terms), or only through $u(k_1|M,z)$ (1h1g term) that we approximate as 1 as argued previously. In that case, the integral over $k$ in Eq. \ref{Eq:crosscov-proj-2redshifts} can be performed analytically and yields:
\ba
\Cov\left(N_\mr{cl}(i_M,i_z) , C_\ell^\mr{gal}(j_z)\right) &= \delta_{i_z,j_z} \int \dd M_1 \, \dd V \, \left.\frac{\dd n_h}{\dd M}\right|_{M_1,z} \! \frac{\nbargal(z)^2}{\Delta N_\mr{gal}(j_z)^2} \, \frac{1}{4\pi} \, B_\mr{hgg}(k,k_\ell,k_\ell | M_1,z)
\ea
This equation allows to compute the cross-covariance from the 2h-2h, 1h2g and 1h1g bispectrum terms (Eq. \ref{Eq:B2h2h}-\ref{Eq:B1h2g}-\ref{Eq:B1h1g}), giving:
\ba
\Cov_\mr{1h1g}\left(N_\mr{cl}(i_M,i_z) , C_\ell^\mr{gal}(j_z)\right) &= \frac{\delta_{i_z,j_z}}{4\pi \; \Delta N_\mr{gal}(j_z)^2} \int \dd V \ n_\mr{gal}(z | i_M) \\
\Cov_\mr{1h2g}\left(N_\mr{cl}(i_M,i_z) , C_\ell^\mr{gal}(j_z)\right) &= \frac{\delta_{i_z,j_z}}{4\pi \; \Delta N_\mr{gal}(j_z)^2} \int \dd V \nbargal(z)^2 \ P_\mr{gal}^\mr{1h}(k_\ell,z | i_M) \\
\nonumber \Cov_\mr{2h-2h}\left(N_\mr{cl}(i_M,i_z) , C_\ell^\mr{gal}(j_z)\right) &= \frac{\delta_{i_z,j_z}}{4\pi \; \Delta N_\mr{gal}(j_z)^2} \int \dd V \ 2 \, \nbargal(z)^2 \, b_1^\mr{gal,eff}(k_\ell,z) \\
& \qquad \times b_1^\mr{gal,eff}(k_\ell,z | i_M) \, P_\mr{DM}(k_\ell,z)
\ea
with
\ba
n_\mr{gal}(z | i_M) &= \int_{M\in\mr{bin}(i_M)} \!\!\!\!\!\!\dd M \ \frac{\dd n_h}{\dd M} \ \lbra N_\mr{gal}(M)\rbra \\
P^\mr{1h}_\mr{gal}(k, z | i_M) &= \frac{1}{\nbargal^2(z)} \int_{M\in\mr{bin}(i_M)} \!\!\!\!\!\!\dd M \ \frac{\dd n_h}{\dd M} \ \lbra N_\mr{gal}(N_\mr{gal}-1)(M)\rbra \ u^2(k|M,z)\\
b_1^\mr{gal,eff}(k, z | i_M) &= \frac{1}{\nbargal(z)} \int_{M\in\mr{bin}(i_M)} \!\!\!\!\!\!\dd M \ \frac{\dd n_h}{\dd M} \ \lbra N_\mr{gal}(M)\rbra \ b_1(M,z) \ u(k|M,z)
\ea
being the contribution of the $i_M$ mass bin to the number of galaxy, 1-halo power spectrum and effective galaxy bias.

Now are left only the 2PT and s2 bispectrum terms which both depend on the angle between the Fourier modes and are thus the subject of the next subsection.

\subsection{Angle dependent bispectrum}\label{Sect:angle-dep-bisp}

As a foreword, one could think that the bispectrum could be parametrised only by the moduli $k_1,k_2,k_3$. Indeed the angles between the vectors can be computed from the moduli, given that $(\kk_1,\kk_2,\kk_3)$ form a triangle. This proves not to be a safe bet, because angles are fast varying functions of the moduli in the squeezed limit $k_1 \ll k_2,k_3$, so we would never be in the position to apply the Limber's approximation used previously. Thus in the following, we keep the angle-dependence of the bispectrum in the equations.

The 2PT and s2 bispectrum terms are of the general form:
\ba
\nonumber B_\mr{halo}(k_{123}|M_{123},z_{123}) &= b_\alpha(M_1,z_1) \, b_1(M_2,z_2) \, b_1(M_3,z_3) \ \mathcal{K}(\kk_2,\kk_3) \ P_\mr{DM}(k_2 | z_{12}) \, P_\mr{DM}(k_3 | z_{13})\\
& + 2 \ \mr{perm.}
\ea
with $\alpha=1,\mr{s2}$ and $\mathcal{K}=2 F^\mr{2PT},S_2$ for the 2PT and s2 terms respectively.\\
As already noted previously for the b2 term, $P(k_2) P(k_3) \ll P(k_1) P(k_2) , P(k_1) P(k_3)$ so we can neglect the corresponding term in the permutation.

The dimensionless kernel $\mathcal{K}$ is a function of the angle between the Fourier modes and can be decomposed onto Legendre polynomials:
\be
\mathcal{K}(\kk_1,\kk_2) = \sum_{\ell_k} \mathcal{K}_{\ell_k} \ P_{\ell_k}(\cos \theta_{12}) 
\ee
For example for the $F^{2PT}$ kernel the sum runs over $\ell_F=0,1,2$ and we have
\be
F^{2PT}_0 = \frac{17}{21} \qquad F^{2PT}_1 = \frac{1}{2} \left(\frac{k_1}{k_2} + \frac{k_2}{k_1}\right) \qquad F^{2PT}_2 = \frac{2}{7}
\ee
while for the $S_2$ kernel, the sum runs only over $\ell_S=2$, all other terms being zero, and we have
\be
S_{2,2} = 1
\ee

Now we need to compute the effect of the $\mathcal{K}_{\ell_k}$ term. Before diving into this, let us first note that the case of $\ell_K=0$ is in fact simple : as $P_0=1$, we are in fact back to the angle-independent case. Through Sect. \ref{Sect:angle-indep-bisp} we know how to tackle this, and we get:
\ba
\nonumber \Cov_\mr{\mathcal{K}_{0}}\left(N_\mr{cl}(i_M,i_z) , C_\ell^\mr{gal}(j_z)\right) &= \frac{2}{\Delta N_\mr{gal}(j_z)^2} \int \dd V_{12} \; \frac{\partial n_h}{\partial \delta_b}(i_M,z_1) \; \nbargal(z_2)^2\\
& \times \mathcal{K}_{0} \, b_{\alpha}^\mr{gal,eff}(k_\ell,z_2) \, b_1^\mr{gal,eff}(k_\ell,z_2) P_\mr{DM}(k_\ell | z_2) \; \sigma^2_\mr{proj}(z_1,z_2) 
\ea
For the s2 term $\mathcal{K}_{0}=0$ while for the 2PT term $\mathcal{K}_{0}=2\times \frac{17}{21}$. Thus:
\ba
\Cov_\mr{s2,0}\left(N_\mr{cl}(i_M,i_z) , C_\ell^\mr{gal}(j_z)\right) &= 0 \\
\nonumber \Cov_\mr{2PT,0}\left(N_\mr{cl}(i_M,i_z) , C_\ell^\mr{gal}(j_z)\right) &= \frac{1}{\Delta N_\mr{gal}(j_z)^2} \int \dd V_{12} \; \frac{\partial n_h}{\partial \delta_b}(i_M,z_1) \; \nbargal(z_2)^2\\
& \times \frac{68}{21} \, b_{1}^\mr{gal,eff}(k_\ell,z_2) \, b_1^\mr{gal,eff}(k_\ell,z_2) P_\mr{DM}(k_\ell | z_2) \; \sigma^2_\mr{proj}(z_1,z_2)
\ea
which correspond to the results stated in the main text for the s2 and 2PT terms.

The rest of this subsection is devoted to argue that the contribution of higher $\ell_K$ is negligible (compared to the 2PT,0 term).\\
To do this we need to go back to the full generality of Eq. \ref{Eq:b0ll-hgg-fullexpanded} for the angular bispectrum :
\ba
\nonumber b^{\mathcal{K}_{\ell_k}}_{0\ell\ell}(M_1,z_{123}) &= \int \frac{\dd^3\kk_{123}}{(2\pi)^9} \, j_0(k_1 r_1) \, j_\ell(k_2 r_2) \, j_\ell(k_3 r_3) \frac{4\pi}{2\ell+1} \sum_m Y^*_{\ell m}(\hk_2) \, Y_{\ell m}(\hk_3) \\
\nonumber & \times B_\mr{hgg}^{\mathcal{K}_{\ell_k}}(\kk_{123}|M_1,z_{123}) \int \dd^3 \xx \, (4\pi)^3 \sum_{1,2,3} i^{\ell_1+\ell_2-\ell_3} \, j_{\ell_1}(k_1 x) j_{\ell_2}(k_2 x) j_{\ell_3}(k_3 x) \\
& \qquad \times Y_1(\hk_1) \, Y_2(\hk_2) \, Y^*_3(\hk_3) \ Y^*_1(\hat{x}) \, Y^*_2(\hat{x}) \, Y_3(\hat{x})
\ea
Now we will drop integrals over masses in the bispectrum, as the argument is based on the integrals of spherical harmonics and spherical bessel function, and the equations are already long enough. We also drop the permutation $2\leftrightarrow 3$ and some redshift dependences for clarity.\\
Basically the angular bispectrum contains:
\ba
\nonumber b^{\mathcal{K}_{\ell_k}}_{0\ell\ell}(M_1,z_{123}) &\ni \int \frac{\dd^3\kk_{123}}{(2\pi)^9} \, j_0(k_1 r_1) \, j_\ell(k_2 r_2) \, j_\ell(k_3 r_3) \frac{4\pi}{2\ell+1} \sum_m Y^*_{\ell m}(\hk_2) \, Y_{\ell m}(\hk_3) \\
\nonumber & \times P_{\ell_K}(\cos\theta_{12}) \, P(k_1) \, P(k_2) \int \dd^3 \xx \, (4\pi)^3 \sum_{1,2,3} i^{\ell_1+\ell_2-\ell_3} \, j_{\ell_1}(k_1 x) j_{\ell_2}(k_2 x) j_{\ell_3}(k_3 x) \\
& \qquad \times Y_1(\hk_1) \, Y_2(\hk_2) \, Y^*_3(\hk_3) \ Y^*_1(\hat{x}) \, Y^*_2(\hat{x}) \, Y_3(\hat{x})
\ea
Now we can expand the Legendre polynomial onto spherical harmonics:
\be
P_{\ell_K}(\cos\theta_{12}) = \frac{4\pi}{2\ell_K +1} \sum_{m_K} Y^*_{\ell_K,m_K}(\hk_1) \, Y_{\ell_K,m_K}(\hk_2)
\ee
Then we can perform the integrals over $\hk_1,\hk_2,\hk_3$. First the integral over $\hk_3$ forces $(\ell_3,m_3)=(\ell,m)$ by orthogonality of spherical harmonics. In the same way, the integral over $\hk_1$ forces $(\ell_1,m_1)=(\ell_K,m_K)$. Finally the integral over $\hk_2$ is non-trivial and yields a Gaunt coefficient:
\be
G(\ell,m,\ell_K,m_K,\ell_2,m_2) = \int \dd^2\hk_2 \; Y^*_{\ell m}(\hk_2) \, Y_{\ell_K,m_K}(\hk_2) \, Y_{\ell_2,m_2}(\hk_2)
\ee
This forces $-m+m_K+m_2=0$ and $|\ell-\ell_K| \leq \ell_2 \leq \ell+\ell_K$ with $\ell+\ell_K+\ell_2$ even.\\
The integral over $\hat{x}$ yields the same Gaunt coefficient (more precisely its complex conjugate, but Gaunt coefficients are real anyway). Thus:
\ba
\nonumber b^{\mathcal{K}_{\ell_k}}_{0\ell\ell}(M_1,z_{123}) &\ni \frac{(4\pi)^5}{(2\pi)^9}\frac{1}{(2\ell+1)(2\ell_K+1)} \sum_{2,m,m_K}\int k^2_{123} \, \dd k_{123} \, x^2 \, \dd x \, j_0(k_1 r_1) \, j_\ell(k_2 r_2) \, j_\ell(k_3 r_3) \\
\nonumber & \times P(k_1) \, P(k_2) \, i^{\ell_K+\ell_2-\ell} \, j_{\ell_K}(k_1 x) j_{\ell_2}(k_2 x) j_{\ell}(k_3 x) G^2(\ell,m,\ell_K,m_K,\ell_2,m_2) \\
\ea
The sum over $m_2,m,m_K$ can be carried out analytically and yields
\be
\sum_{m_2,m,m_K} G^2(\ell,m,\ell_K,m_K,\ell_2,m_2) = \frac{(2\ell_2+1)(2\ell+1)(2\ell_K+1)}{4\pi} \; 
\begin{pmatrix}
\ell & \ell_2 & \ell_K \\ 0 & 0 & 0
\end{pmatrix}^{2}
\ee
The integral over $k_3$ is the already-used identity of integral of spherical Bessel functions and yields a dirac of redshifts $z_3$ and $z_x$.

Now comes the pitch. For the cases of interest (2PT and s2 bispectrum terms), $\ell_K=1$ or 2. Thus, $\ell_2=\ell-1,\ell+1$ or $\ell_2=\ell-2,\ell,\ell+2$ respectively. Let's compare the $k_1$ and $k_2$ integrals to the $\ell_K=0$ case. In the $\ell_K=0$ case we have
\ba
\int k^2_1 \, \dd k_1 \; j_0(k_1 r_1) \, j_0(k_1 x) \; P(k_1) \\
\int k^2_2 \, \dd k_2 \; j_\ell(k_2 r_2) \, j_\ell(k_2 x) \; P(k_2)
\ea
while for a general $\ell_K$ we have
\ba
\int k^2_1 \, \dd k_1 \; j_0(k_1 r_1) \, j_{\ell_K}(k_1 x) \; P(k_1) \\
\int k^2_2 \, \dd k_2 \; j_\ell(k_2 r_2) \, j_{\ell_2}(k_2 x) \; P(k_2)
\ea
We expect both integrals to be small compared to their relative counterparts in the $\ell_K=0$ case. Indeed, the Bessel functions will have destructing interferences. They will not have the same frequency, unless $z_1=z_x$ for the first integral ($z_2=z_x$ for the second) where the integrals of the $\ell_K=0$ case are maximal. And in the $z_1=z_x$ (resp. $z_2=z_x$) case, the Bessel interferences will still be destructive because:
\begin{itemize}
\item $j_0$ and $j_{\ell_K}$ do not peak at the same location. And the same is true for $j_\ell$ and $j_{\ell_2}$.
\item the asymptotic oscillations\footnote{$j_n(y) \sim \sin(y-\frac{n\pi}{2})/y$ for $y\gg n$.} are out of phase.
\end{itemize}

We thus conclude that these terms are negligible compared to the $\ell_K=0$ case. Thus (2PT,2)$\ll$(2PT,0) and (2PT,1)$\ll$(2PT,0). Given that the $F^\mr{2PT}$ and $S_2$ kernels are both of order 1 and that the corresponding biases will also be of the same order\footnote{In fact $b_\mr{s2}=-\frac{2}{7}(b_1-1)$ following \cite{Baldauf2012}.}, we also conclude that (s2,2)$\ll$(2PT,0).

\section{SSC inequalities}\label{App:SSC-ineq}
This appendix contains the mathematical proof of the SSC inequalities stated in the main text, as well as a discussion of their consequences.\\

\noindent {\bf Definition:} A bivariate symmetric function $\kappa(x,y)$ is said positively separable if it is of the form
$$ \kappa(x,y) = \sum_i \Psi_i(x) \, \Psi_i(y) $$
for some family of univariate functions $\Psi_i$.\\

\noindent {\bf Lemma:} If $\kappa(x,y)$ is positively separable, the bivariate form
$$ <f,g> = \int \dd x \, \dd y \; f(x) \, g(y) \; \kappa(x,y)$$
is a semi-inner product over real-valued functions.\\

\noindent \textit{Proof:}\\
The form is obviously bilinear, and symmetric because of the symmetry of $\kappa(x,y)$.\\
Let's prove positivity, i.e. $<f,f> \geq 0$:
\ba
\nonumber <f,f> &= \int \dd x \, \dd y \; f(x) \, f(y) \; \kappa(x,y)\\
\nonumber &= \sum_i \int \dd x \, \dd y \; f(x) \, f(y) \; \Psi_i(x) \, \Psi_i(y)\\
&= \sum_i \left(\int f \, \Psi_i \right)^2 \\
\nonumber & \geq 0
\ea

\noindent \textit{Note:} $<f,f> = 0 \Leftrightarrow \int f \, \Psi_i$ $\forall i$. Thus we have positive definiteness, i.e. $<f,g>$ is a strict inner product, iff the family $(\Psi_i)$ is a basis of real-valued functions.
\newline

\noindent {\bf Consequence:} We have the Cauchy-Schwarz inequality \cite{Schwarz1888}
$$ |<f,g>| \leq \| f \| \cdot \| g\|$$

\noindent {\bf Application:} Recall the equations for super-sample covariance of $C_\ell^\mr{gal}(i_z)$ and $N_\mr{cl}(i_M,i_z)$:
\ba
\Cov_\mr{SSC}\left(N_\mr{cl}(i_M,i_z) , N_\mr{cl}(j_M,i_z))\right) &= \int \dd V_{12} \; \frac{\partial n_h(i_M,z_1)}{\partial \delta_\mr{b}} \, \frac{\partial n_h(j_M,z_2)}{\partial \delta_\mr{b}} \; \sigma^2_\mr{proj}(z_1,z_2)  \\
\Cov_\mr{SSC}\left(N_\mr{cl}(i_M,i_z) , C_{\ell}^\mr{gal}(j_z)\right) &= \int \dd V_{12} \; \frac{\nbargal(z_2)^2}{\Delta N_\mr{gal}(j_z)^2} \; \frac{\partial n_h(i_M,z_1)}{\partial \delta_\mr{b}} \, \frac{\partial P_\mr{gal}(k_{\ell},z_2)}{\partial \delta_\mr{b}} \; \sigma^2_\mr{proj}(z_1,z_2)  \\
\Cov_\mr{SSC}\left(C_\ell^\mr{gal}(i_z) , C_{\ell'}^\mr{gal}(j_z)\right) &= \int \dd V_{12} \; \frac{\nbargal(z_1)^2 \, \nbargal(z_2)^2}{\Delta N_\mr{gal}(i_z)^2 \, \Delta N_\mr{gal}(j_z)^2} \; \frac{\partial P_\mr{gal}(k_\ell,z_1)}{\partial \delta_\mr{b}} \, \frac{\partial P_\mr{gal}(k_{\ell'},z_2)}{\partial \delta_\mr{b}} \; \sigma^2_\mr{proj}(z_1,z_2)
\ea
where $\dd V = r^2(z) \frac{\dd r}{\dd z} \, \dd z$, the redshift integrals are implicitly over the corresponding redshift bins, and
$$ \sigma^2_\mr{proj}(z_1,z_2) = \frac{1}{2\pi^2} \int k^2 \dd k \; j_0(k r_1) \, j_0(k r_2) \; P(k| z_{12})$$
Given that $P(k| z_{12}) = G(z_1) \, G(z_2) P(k,z=0)$ with $G(z)$ the growth function, $\sigma^2_\mr{proj}(z_1,z_2)$ is positively separable with
$$ \sum_i \rightarrow \frac{1}{2\pi^2} \int k^2 \dd k \quad \mr{and} \quad \Psi_i \rightarrow j_0\left(k r(z)\right) \, G(z) \, \left(P(k,z=0)\right)^{1/2}$$
We are thus in position to apply the lemma, with e.g.
$$ f(z) = r^2(z) \frac{\dd r}{\dd z} \; \frac{\partial n_h(i_M,z)}{\partial \delta_\mr{b}} \; \mathds{1}_{[z \in \mr{bin}(i_z)]} \quad \mr{and} \quad g(z) = r^2(z) \frac{\dd r}{\dd z} \; \frac{\partial n_h(j_M,z)}{\partial \delta_\mr{b}} \; \mathds{1}_{[z \in \mr{bin}(j_z)]}$$
giving
\be
\left|\Cov_\mr{SSC}\left(N_\mr{cl}(i_M,i_z) , N_\mr{cl}(j_M,j_z))\right)\right| \leq \left[\Var_\mr{SSC}\left(N_\mr{cl}(i_M,i_z)\right) \cdot \Var_\mr{SSC}\left(N_\mr{cl}(j_M,j_z)\right) \right]^{1/2}
\ee
In particular, for a given redshift slice $i_z=j_z$, this inequality ensures that the correlation matrix of cluster counts has off-diagonal elements $\leq 1$ even in the deep SSC domination regime.\\
The same process can be applied to the auto-covariance of the galaxy power spectrum, yielding:
\be
\left|\Cov_\mr{SSC}\left(C_\ell^\mr{gal}(i_z) , C_{\ell'}^\mr{gal}(j_z))\right)\right| \leq \left[\Var_\mr{SSC}\left(C_\ell^\mr{gal}(i_z)\right) \cdot \Var_\mr{SSC}\left(C_{\ell'}^\mr{gal}(j_z)\right) \right]^{1/2}
\ee
Again ensuring that the auto correlation matrix of the galaxy spectrum has off-diagonal elements $\leq 1$ in the SSC dominated regime.\\
Finally we can apply the Cauchy-Schwarz inequality to the cross-covariance, with
$$ f(z) = r^2(z) \frac{\dd r}{\dd z} \; \frac{\partial n_h(i_M,z)}{\partial \delta_\mr{b}} \; \mathds{1}_{[z \in \mr{bin}(i_z)]} \quad \mr{and} \quad g(z) = r^2(z) \frac{\dd r}{\dd z} \; \frac{\nbargal(z)^2}{\Delta N_\mr{gal}(j_z)^2} \frac{\partial P_\mr{gal}(k_\ell,z)}{\partial \delta_\mr{b}} \; \mathds{1}_{[z \in \mr{bin}(j_z)]}$$
yielding
\be
\left|\Cov_\mr{SSC}\left(N_\mr{cl}(i_M,i_z) , C_\ell^\mr{gal}(j_z))\right)\right| \leq \left[\Var_\mr{SSC}\left(N_\mr{cl}(i_M,i_z)\right) \cdot \Var_\mr{SSC}\left(C_{\ell}^\mr{gal}(j_z)\right) \right]^{1/2}
\ee
ensuring that the normalised joint correlation matrix has off-diagonal elements $\leq 1$.\\
\newline
These inequalities are most useful when $i_z=j_z$. Indeed the kernel $\sigma^2_\mr{proj}(z_1,z_2)$ decreases quickly when $z_1$ becomes different from $z_2$, thus the SSC covariance decreases quickly for different redshift bins.
\newline

\noindent{\bf Consequence 2:} The SSC joint covariance is a positive matrix.
\newline

\noindent\textit{Note:} This statement is in fact more general and implies in particular the Cauchy-Schwarz inequality
\newline

\noindent\textit{Proof:} The proof simply amounts to showing that the covariance $C$ is a semi-inner product.\\
We have $C_{ij} = \int \dd V_{12} \; \frac{\partial O_i}{\partial \delta_\mr{b}} \, \frac{\partial O_j}{\partial \delta_\mr{b}} \; \sigma^2_\mr{proj}(z_1,z_2) $\\
with observable i $O_i$ being in our case either $n_h(i_M,z)$ or $\frac{\nbargal(z)^2}{\Delta N_\mr{gal}(j_z)^2} \; P_\mr{gal}(k_\ell,z)$.\\
Thus:
\ba
\phantom{X}^\mr{T}X \cdot C \cdot X &= \sum_{ij} \int \dd V_{12} \; k^2 \dd k \; j_0(k r_1) \, j_0(k r_2) \; P(k| z_{12}) \; x_i \frac{\partial O_i}{\partial \delta_\mr{b}} \, \frac{\partial O_j}{\partial \delta_\mr{b}} x_j \\
&= \int k^2 \dd k \left[\int \dd V \, j_0(k r) \, G(z) \,  \big(P(k,z=0)\big)^{1/2} \sum_i x_i \, \frac{\partial O_i}{\partial \delta_\mr{b}}\right]^2\\
\label{Eq:C_SSC_>=0} & \geq 0
\ea
So that $C$ defines a semi-inner product.\\
\newline

{\bf Is $C$ definite ?}\\
To investigate definiteness, we have to examine the case of equality in Eq. \ref{Eq:C_SSC_>=0}.
\ba
\phantom{X}^\mr{T}X \cdot C \cdot X = 0 \quad &\Leftrightarrow \quad \forall k \quad \int \dd V \, j_0(k r) \, G(z) \, \sum_i x_i \, \frac{\partial O_i}{\partial \delta_\mr{b}} = 0
\ea
$j_0(k r) = \sin(kr)/kr$ so we have basically $\int \dd r \sin(kr) \; f(r) =0$ $\forall k$ so $f(r)=0$ (except an unphysical dirac at r=0). Thus
\ba
\phantom{X}^\mr{T}X \cdot C \cdot X = 0 \quad &\Leftrightarrow \quad \forall z \quad G(z) \, \sum_i x_i \, \frac{\partial O_i}{\partial \delta_\mr{b}} = 0 \\
&\Leftrightarrow \quad \forall z \quad \frac{\partial \sum_i x_i \, O_i}{\partial \delta_\mr{b}} = 0
\ea
So the covariance is not definite \textit{iff} there is some linear combination of observables which does not react to any change of background at any redshift.\\
This would happen for instance if we included twice the same observable in the data vector, which we obviously do not do, or if two observables reacted exactly in the same way to a background change, in which case we would see a 100\% SSC correlation for them. From the figures given in the text, this is not the case of the cluster counts ; however it does appear to be the case for the galaxy power spectrum, to the limit of numerical precision. Hence $C$ may have some zero eigenvalues, or at least eigenvalues small enough to blow up its condition number. Numerical regularization, e.g. with Principal Component Analysis (PCA), would thus be required to analyse observables deep in the SSC dominated regime.

\section{Multipole binning}\label{App:bin-multipoles}
A binning scheme can be defined by its stakes $\ell_b$, such that the bin with index $b$ contains the multipoles $\ell_b \leq \ell < \ell_{b+1}$, with the center of the bin noted as $\ell_\mr{cent}$ and its width as $\Delta\ell$. A weighting is usually applied to the power spectrum to make the binned elements closer to constant so as to lose less information :
\ba
D_\ell &= w_\ell \; C_\ell \\
D_b &= \frac{1}{\Delta\ell}\sum_{\ell \in [ \ell_b,\ell_{b+1} [} D_\ell
\ea
where the weight $w_\ell$ is usually $\frac{\ell(\ell+1)}{2\pi}$ in CMB studies. In our case, the galaxy power spectrum behaves approximately as $C_\ell \propto 1/ell$ so that a weight $w_\ell = \ell$ is more adapted.\\
The binned $D_b$ can optionally be transformed back into a spectrum-like quantity with $C_b = D_b / w_{\ell_\mr{cent}}$. The information content in $D_b$ and $C_b$ is the same but the latter is useful for visualization purpose when performing actual data analysis.\\
We can thus define the binning operator
\be
P_{b,\ell} = 
\left\{
\begin{array}{ll}
\frac{w_\ell}{w_{\ell_\mr{cent}} \, \Delta\ell} & \qquad \mathrm{if} \quad \ell \in [ \ell_b,\ell_{b+1} [ \\
0 & \qquad \mathrm{otherwise} \\
\end{array}
\right.
\ee
so that
\ba
C_b &= P_{b,\ell} \; C_\ell \\
\Cov(C_b,C_{b'}) &= P_{b,\ell} \; P_{b',\ell'} \; \mathcal{C}_{\ell,\ell'}
\ea

To apply these equations in our case, we need predictions of the power spectrum and more difficultly of its covariance for all multipoles in the range considered. Exact analytical prediction however proves too numerically-intensive to be achieved. Nevertheless the power spectrum and its covariance prove remarkably smooth when considered in the adequate space. Thus we employ interpolation. More precisely we use spline interpolation in the space $(\log\ell,\log C_\ell)$, which works very accurately for the power spectrum. For the covariance, there are two small problems to overcome : the covariance can be negative, and the Gaussian term is a sharp peak on the diagonal. The sign of the covariance is fortunately constant with multipoles for a given couple of redshift bin, so that we can factor out the sign and interpolate $\log \left|\mathcal{C}_{\ell,\ell'}\right|$. The diagonal of the Gaussian covariance is then interpolated separately from the other terms which have a smooth off-diagonal behaviour.

\begin{figure}[htb]
\centering
\includegraphics[width=.3\linewidth]{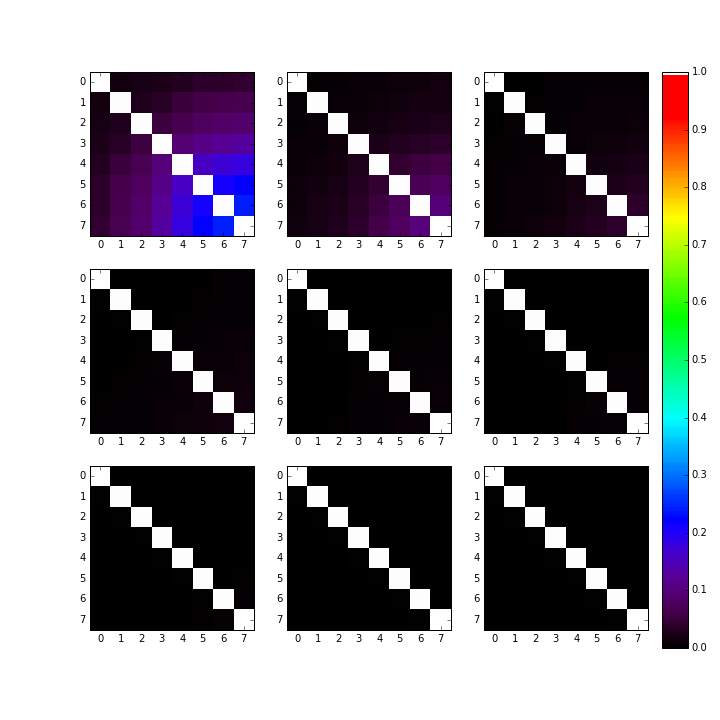}
\includegraphics[width=.3\linewidth]{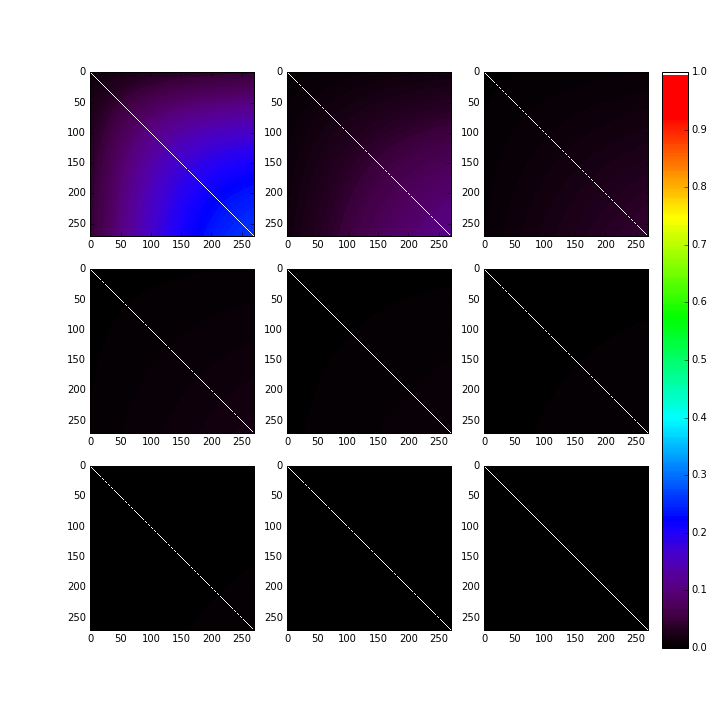}
\includegraphics[width=.3\linewidth]{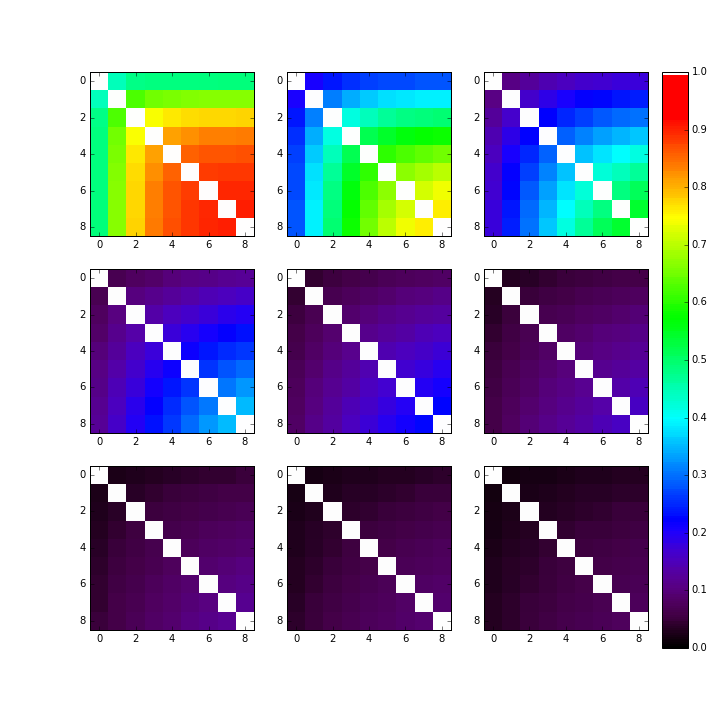}
\caption{Correlation matrix of the galaxy power spectrum in nine redshift bins. \textit{Left:} Correlation matrix of eight particular multipoles between $\ell_\mr{min}=30$ and $\ell_\mr{max}=300$. \textit{Center:} after interpolation, correlation matrix of all multipoles in $30\leq\ell\leq 300$. \textit{Right:} correlation matrix after binning of multipoles with a constant width $\Delta\ell=30$.}
\label{Fig:comp-covcl_raw-interp-binned}
\end{figure}

Figure \ref{Fig:comp-covcl_raw-interp-binned} shows the results of the different steps : first the covariance on the original eight multipoles computed, then the interpolated covariance on the full multipole range, and finally the binned covariance matrix. 

One sees that binning increases the off-diagonality of the matrix, and makes thus more evident the importance of non-Gaussianity on the information content of the galaxy power spectrum. Indeed, the Gaussian covariance being diagonal, its contribution decreases with the bin width as $\Var(C_b) \sim \Var(C_{\ell_{cent}}) / \Delta\ell$, while non-Gaussian terms decrease much more slowly. For example in the case of super-sample covariance, multipoles are nearly 100\% correlated together if SSC is the only covariance contributor, meaning the correlation matrix has value 1 everywhere. After binning, the bins remain obviously nearly 100\% correlated together so that the binned SSC correlation matrix is as off-diagonal as the unbinned one.

We note that this result might be seen as counter-intuitive at first sight, since averaging should Gaussianize a variable by the central limit theorem. But the paradox is only at first sight : the central limit theorem ensures that the \emph{likelihood} of $C_b$ becomes closer to Gaussian as $\Delta\ell$ increases, but it tells nothing about the evolution of the covariance matrix of $C_b$. Physically, the constant effect of mode-coupling (due to non-linearity) becomes more evident as the error bars on mode determination shrink (via binning).

\acknowledgments
This work was supported by the S\~ao Paulo Research Foundation (FAPESP)
under grants \#2011/11973 and \#2013/19936-6. 
R.R.~was partially supported by a CNPq research grant.
The authors acknowledge useful discussions with Michel Aguena, Scott Dodelson, Enrique Gazta\~{n}aga, Kai Hoffmann, Benjamin Joachimi and Elisabeth Krause.





\end{document}